\documentclass[sigconf]{acmart}

\usepackage{tabularx}
\usepackage{graphicx}
\usepackage{subfigure}
\usepackage{threeparttable}
\usepackage{longtable}
\usepackage{multicol}
\usepackage{multirow}
\usepackage{pifont}
\usepackage{url}
\usepackage{hyperref}

\usepackage{makecell}
\usepackage{soul} 
\usepackage{color, xcolor} 

\AtBeginDocument{%
  \providecommand\BibTeX{{%
    \normalfont B\kern-0.5em{\scshape i\kern-0.25em b}\kern-0.8em\TeX}}}

\setcopyright{acmcopyright}
\copyrightyear{2022}
\acmYear{2022}
\acmDOI{10.1145...}

\acmConference[XXXX]{XXXXX}{XXXXX}{XXXXX}
\acmPrice{XXXX}
\acmISBN{XXXXXX}

\begin{document}

\title{ShieldMMU: Detecting and Defending against Controlled-Channel Attacks in Shielding Memory System}

\author{Gang Liu}
\email{gliu29@uestc.edu.cn}
\orcid{0000-0003-1330-0945}
\affiliation{%
  \institution{Shenzhen Institute for Advanced Study, University of Electronic Science and Technology of China}
  \streetaddress{Phase II of Yinxing Zhijie, Guanlan Street, Longhua District}
  \city{Shenzhen}
  \state{Guangdong}
  \country{China}
  \postcode{518110} 
}  
 \author{Ningjie Li}
  \email{njli@shu.edu.cn}
  \affiliation{%
  \institution{School of Computer Engineering and Science, Shanghai University}
  \streetaddress{No. 99 Shangda Road, Baoshan District}
  \city{Shanghai}
  \country{China}
  \postcode{200444}
  }

   \author{Cen Chen}
  \email{chencen@scut.edu.cn}
  \affiliation{%
  \institution{School of Future Technology, South China University of Technology}
  \streetaddress{Room B1-c419, Guangzhou International Campus}
  \city{Guangzhou}
  \state{Guangdong}
  \country{China}
  \postcode{510641}
  }

\renewcommand{\shortauthors}{Gang Liu, et al.}

\begin{abstract}
Intel SGX and hypervisors isolate non-privileged programs from other software, ensuring confidentiality and integrity. However, side-channel attacks continue to threaten Intel SGX's security, enabling malicious OS to manipulate PTE present bits, induce page faults, and steal memory access traces. Despite extensive research, existing defenses focus on detection or rely on impractical solutions.
This paper presents ShieldMMU, a comprehensive solution for mitigating controlled channel attacks, balancing compatibility, performance, and usability. Leveraging a Merkle Tree-inspired Defense Tree (DD-Tree), ShieldMMU protects PTE integrity by detecting, locating, and restoring attacked PTEs. It identifies MMU page table lookup events and side-channel attacks, promptly restoring PTE parameters to prevent page fault traps and ensure secure non-privileged application operation within SGX.
Our experiments confirm ShieldMMU's enhanced security and acceptable latency performance.
\end{abstract}





\maketitle

\section{Introduction}
{The persistent iterative development of contemporary general-purpose OS and applications has resulted in codebases that readily scale to millions of lines of code, where a singular vulnerability can frequently precipitate a complete breach of all security assurances.
To mitigate the aforementioned challenges, both academic and industrial sectors have proposed and advanced Trusted Execution Environment (TEE) technologies \cite{Alves_2004, McKeen2013}, which implement an alternative, non-hierarchical security paradigm for isolated application domains known as enclaves.
With Intel's release of Software Guard Extensions (SGX) \cite{mckeen2013innovative, anati2013innovative}, which provides hardware-enforced TEE isolation and attestation guarantees, coupled with its promise of robust security assurances, an increasing number of applications in the industry are adopting this technology to ensure the security of programs executed on shared machines.
Intel SGX is designed with hardware-based functionalities, aiming to allow applications to achieve self-protection when running on an OS, thereby mitigating attacks from the Operating System (OS), hypervisors, BIOS, and other software.
}

{While hardware-based Intel SGX provides strong security guarantees for programs and data running within enclaves, research has demonstrated that this technology remains vulnerable to threats such as page fault-based and cache side-channel attacks \cite{shinde2016preventing, Controlled-Channel-Attacks}.
The core premise of these methods \cite{shinde2016preventing, Controlled-Channel-Attacks} assumes the OS is compromised or untrusted. 
Attackers leverage the OS to restrict physical page allocations for sensitive code or data, forcing the application's memory accesses to spill beyond the allocated physical pages, thereby triggering page faults. 
By monitoring these faults and statistically analyzing the application's access patterns, attackers infer critical information about secret-dependent memory operations.
Page fault side-channel attacks (also referred to as controlled-channel attacks) rooted in a malicious OS can circumvent the security protections of SGX technology, as the adversarial OS retains complete control over the page access behavior of SGX-enclaved processes.
}
{In real-world application scenarios, untrusted OS still provides critical functionalities such as memory resource management and standardized public interfaces. 
Consequently, an untrusted OS can subvert shielded execution environments by maliciously manipulating system call return values (e.g., via IAGO attacks \cite{Iago}), exploiting its privileged control over kernel-space operations to bypass hardware-enforced isolation guarantees.
}

{Several research efforts have been proposed to mitigate the threat of malicious OS exploiting the page fault channel to leak sensitive information \cite{shinde2016preventing, chen2017detecting, shih2017t, liu2022ps, orenbach2020autarky}.
Shinde et al \cite{shinde2016preventing}. proposed a security property termed "page-fault obliviousness" to safeguard applications against information leakage through the page-fault channel exploited by a malicious OS. 
Programmers are required to manually annotate the portions of the code that depend on sensitive input.
The compiler then performs a static analysis based on these annotations and applies the corresponding code transformations to ensure that the page access patterns remain consistent during execution, regardless of the input. 
This achieves PF-obliviousness (page-fault obliviousness), effectively masking input-dependent memory-access behaviors from potential adversaries.
It is not automated and relies on manual and application-specific developer annotations, thus increasing the burden on programmers.
Similar related works include \cite{orenbach2020autarky, shinde2016preventing, wichelmann2024obelix, yadalam2020sgxl}.
Déjà Vu \cite{chen2017detecting} leverages another feature of modern Intel platforms, i.e., Transactional Synchronization Extensions (TSX), which is a hardware implementation of transactional memory. 
This design uses transactional memory to advance its reference clock, making it highly likely that interrupts or page faults will cause the transaction to abort.
The technology detected a page anomaly and consequently halted the application's execution, which is not user-friendly. 
In practice, users desire not only the detection of adversarial attacks but also the ability to deceive such attacks, allowing the application to continue running without compromising privacy and security information.
Memory access obfuscation methods typically incur a high-performance cost (usually more than $2\times$, e.g., \cite{liu2022ps}).
}

{Therefore, the aforementioned related research primarily faces three types of challenges: 
1) lack of automation, 2) low performance, and 3) absence of mechanisms that can effectively circumvent adversarial attacks and continue application execution.
To protect applications from the aforementioned page fault side-channel attacks, we seek a security property that enables applications to detect adversarial attacks and deceive adversaries while running in an untrusted OS environment. This property ensures that the OS cannot obtain any sensitive information by observing page faults. We refer to this property as page fault insensitivity.
}

{In this paper, to address the aforementioned challenges, we propose a novel defense mechanism called \textbf{shieldMMU} that is capable not only of detecting controlled channel attacks within the Memory Management Unit (MMU), but also of enabling applications to bypass attacker assaults and continue operating normally. 
In other words, this defense method can protect the application's normal operation without the attacker's knowledge, thereby rendering conventional page fault attack methods ineffective.
shieldMMU ensures that protected applications perform page address verification during execution through a defense tree called \textbf{DD-Tree}(i.e., capable of both \textbf{D}etecting and \textbf{D}efending against adversary attacks)
Upon successful validation, program execution continues. Otherwise, the true content of the fault page is directly accessed through an address page restoration mechanism.
}

{Specifically, we employ a defense tree similar to a Merkle tree (MT) to safeguard Page Table Entries (PTEs) that operate in memory. When the MMU requires a PTEs lookup during address translation and finds the corresponding PTEs in the page table but with the present bit marked as 0 (a primary indicator of a controlled side-channel attack), the defense tree mechanism is activated to perform data integrity verification on the PTEs. If integrity verification fails, it means that an attacker is executing a side-channel attack. 
Unlike the previous method \cite{chen2017detecting}, our proposed mechanism \textbf{shieldMMU} does not interrupt the application upon integrity verification failure (page fault). Instead, it continues to access the memory page using the physical address protected by the defense leaf node and reverts the present bit to allow unrestricted access. This approach effectively deceives the adversary's attack, thereby avoiding falling into the attacker's trap.
The experimental results demonstrate that shieldMMU is practically feasible, as it is designed based on the underlying hardware system and can be easily deployed on existing Intel SGX processors. Compared to previous controlled-channel defense schemes, shieldMMU does not require programmers to perform special annotation processing during programming, and its performance is acceptable.}

In summary, this paper makes the following contributions:
\begin{itemize}

    
    \item {We analyze and implement controlled side-channel attacks, employing both theoretical and experimental methodologies to characterize the attributes of page fault attacks.}

    
    \item {We propose a defense mechanism called shieldMMU, which allows software within an Intel SGX enclave to defend against controlled side-channel attacks by deploying defenses within the hardware. Our approach aims to detect whether modifications made by a malicious OS are indeed malicious or part of normal operations, thereby serving as a detection mechanism.}

    \item{We introduce the Page Upper Directory (PUD) to decompose the defense tree into a multi-tree defense forest to improve the efficiency of searching and updating defense trees.}

    
    
    \item {We propose a defense mechanism against controlled side-channel attacks. Upon detecting an attack via shieldMMU detection, our method triggers an address translation within the MMU to directly utilize the physical base address protected by a defense tree. Simultaneously, it promptly restores the PTEs, thereby preventing the system from falling into attack-induced traps and achieving effective defense.}

    \item{We implement a controlled side-channel attack within a simulated environment conducive to an attacker, established using the Gem5 simulator. The effectiveness of our approach is evaluated through experimental validation, observation, and statistical inference.}
\end{itemize}

{
The remainder of this paper is organized as follows. Section \ref{Background} provides a details introduction to the Shielded Execution Environment (Intel SGX), MMU, and MT.
Section \ref{controlled-channelAttack} describes controlled-channel attacks in depth.
Section \ref{Relatedwork} surveys the current status of related research work.
Section \ref{design} explains the design.
Section \ref{evaluation} shows our evaluation results. 
}

\section{Hardware Primitives and Motivation} \label{Background}
\subsection{Shielded Systems and Intel SGX}
{
The untrusted OS isolates a portion of space to protect the confidentiality and integrity mechanisms of specific code or data, preventing malicious adversaries from stealing private information. Consequently, this system is referred to as Shielded Systems \cite{lie2003implementing, ta2006splitting,mccune2008flicker, yang2008using, ports2008towards, zhang2011cloudvisor, hofmann2013inktag, cheng2013appshield, criswell2014virtual, baumann2015shielding, chen2017detecting}. Shielded Systems utilize trusted hardware or a hypervisor to prevent the OS from reading or writing to the application's memory.
}

{The advent of Intel SGX \cite{rozas2013intel} enhances the security of shielded systems by protecting enclave memory regions from unauthorized external software reads and writes, regardless of application privilege levels.
Haven system \cite{baumann2015shielding} prevents attacks from malicious OS, such as Iago attacks, by running unmodified applications within an SGX-protected execution environment and utilizing a LibOS to handle interactions with the OS.
}
Consisting of the Windows user-mode library and the user-mode kernel, Drawbridge provides applications with a full Windows interface, but only a very narrow set of dependencies on the underlying system. 
This set of dependencies is significantly smaller than the system call interface for Windows or Linux. The Shield module protects the remaining dependencies of the host OS from adversarial operations. 
For example, the Shield module implements an encryption and integrity-protected file system that restricts storage interaction with untrusted host OS to read and write cryptographically protected disk blocks.
When a page failure occurs, SGX clears the lowest 12 bits of the failed address. 
Control is then transferred to the exception handler. Finally, after completing the process, the handler resumes the Enclave program. 

{In summary, the shielding system incorporates a TEE to ensure the security of application memory pages. Applications executing within the TEE are afforded dual protection of confidentiality and integrity, with the MMU enforcing isolation of user-space program memory regions from other software.}


\subsection{MMU and Page Table}

The MMU is a hardware module used to implement virtual memory management between the CPU and memory, and its main function is to convert virtual addresses into physical addresses.

The mapping relationship between virtual addresses and physical addresses is stored in page tables, while modern processors mostly use hierarchical page tables(PT), such as 64-bit virtual address systems support 48-bit virtual address space and use 4-level page tables. The 48-bit virtual address is split between the Virtual Page Number (VPN) and the page offset. Let's assume 4 KB of data and a PT page. So, the VPN is 36-bit and the page offset is 12-bit. Each PT page consists of 8-byte PTEs, and each PT page yields 512 PTEs. The VPN bits are split into 4 PT indexes (9 bits per PT index), and which PTE is chosen at each PT level. We refer to each level of PT as the PT index of $PL_4$, $PL_{3}$, $PL_2$, $PL_1$, and VPN, respectively, $PLI_4$, $PLI_3$, $PLI_2$, $PLI_1$. During page traversal, the MMU traverses the PT tree starting at $PL_4$. The register $cr3$ points to $PL_4$ and uses the $PLI_4$ index to get $PL_3$ and so on. This process is terminated when the MMU reaches the last PT, where $PLI_1$ returns the PTE given the physical page number for accessing the data page. Page offsets don't need to be translated because they are shared between virtual and physical addresses.

The $0-$th bit of an 8-byte PTEs is the present bit, which indicates whether the next page table or page exists in memory. If there is no next-level page table or page in memory, i.e., the present bit is 0, then the MMU will trigger a page fault, which can occur at any level of the PT tree.

\begin{figure}[!htbp]
\centering
\begin{minipage}{0.45\textwidth} 
    \centering 
    \includegraphics[width=\textwidth]{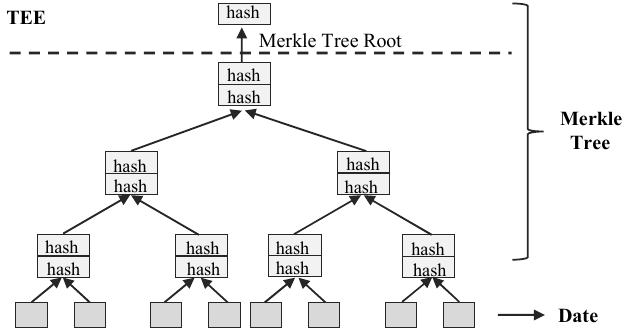} 
    \caption{Merkle Tree} 
    \label{MerkleTreeFig1}
\end{minipage}
\end{figure}

\subsection{Merkle Tree}
The MT \cite{Merkle_1988}\cite{Gassend_Suh_Clarke_van}, also known as a hash tree, is a type of binary tree used to verify and guarantee the integrity and consistency of data. As shown in Fig \ref{MerkleTreeFig1}. MT protects the integrity of data through a set of hashes. Each leaf node of the MT stores the hash value of the data block, and each non-leaf node stores the combined hash value of its child node, which is usually the hash value calculated after the hash value of the two child nodes is concatenated. Follow this process and continue to grow MT until it converges to a node called the root. In general, a tree node can be stored in a non-secure environment, while the root is stored in a trusted secure environment and never leaves it. Each time a data integrity is verified, the entire path from the leaf node to the hash of the root is verified. And because the root never leaves the TEE, it reflects the latest state of the data, provides a basis for tree-wide integrity checks, and is able to detect any tampering with the data. MT has been one of the most commonly used data validation schemes because it has better protection than many others \cite{Rogers_Chhabra_Prvulovic_Solihin_2007}\cite{Berbecaru_Albertalli_2008}.

A variant of theMT, known as the Bonsai Merkle tree (BMT), has been widely used in recent years due to its lower overhead than the original MT scheme \cite{Freij_Zhou_Solihin_2021, Huang_Hua_2021, Elbaz_Champagne_Gebotys_Lee_Potlapally_Torres_2009}. The root of BMT can be updated using eagerly updated or lazy updating techniques, especially if a metadata cache exists \cite{Shadab_Zou_Gandham_Awad_Lin_Mingjie}. Allows a small number of intermediate nodes of the tree to be cached in a secure environment, and in order to reduce the performance overhead associated with tree traversal, both the update and validation processes are stopped on the first node hit in the cache. For updates, when these nodes are evicted from the cache, they propagate the update to their parent nodes. This is known as a lazy update scenario, and updates that always go into the root are part of an eager update scenario.

\subsection{Motivation}

\subsubsection{Theory Motivation} \label{theorymotivation}
{In the management of virtual address space, the address space of each job is divided into blocks that correspond to and match the addresses of physical memory blocks. Each of these blocks is referred to as a page. Physical memory can achieve flexible and discrete page allocation through virtual memory management, thereby reducing the issue of physical memory fragmentation and enhancing the utilization efficiency of physical memory. This mapping from pages to physical memory blocks is known as a page table, which is typically a single-level page table, though in some cases, it may be a multi-level page table, managed by an untrusted OS. When a program needs to access another page, it queries the page table to determine whether the page is in a physical memory block, thereby deciding whether it is necessary to fetch the required block (page) from external memory. Even in Intel SGX, the OS retains the capability to manage the page table, and memory swapping is ultimately performed by the OS.}

{While shielding systems, through trusted hardware or hypervisors, provide confidentiality and integrity to applications by isolating them from untrusted OS, the untrusted OS continue to supply critical functionalities such as memory management and exposed standard interfaces. Many legacy software applications are written under the assumption of a trusted OS, making complete isolation of applications from the OS challenging through shielding systems. Furthermore, the design of shielding systems often overlooks side-channel attacks.}


{In this process, if the OS is malicious, side-channel threats become the primary attack vector targeting the shielded system protector, which is highly susceptible to precise control channel attacks. As proposed in \cite{Controlled-Channel-Attacks}, in control channel attacks, a malicious OS of the shielded system, by retaining privileged behaviors in memory management, can restrict an application's access to specific pages through unrestricted access and manipulation of page tables. This occurs because the application's privacy-related data flows and control flows are exposed to the malicious OS.}


Most existing defense efforts protect data flow and control flow by hiding or obfuscating page access patterns \cite{brasser2019dr, mckeen2016intel, zhang2020klotski, shih2017t, shinde2016preventing}. However, such methods incur significant latency and performance overhead. Another approach involves using a software reference clock to check the execution time of the program to detect the occurrence of attacks \cite{chen2017detecting}. However, this method merely terminates the program upon detecting an attack and does not achieve the goal of defense.
Against adversary attacks, our goal is not only to detect but also to defend at the source, i.e., to continue to execute the application without the attack of the malicious OS, giving the attacker the illusion that the attack has been carried out. Since the main feature of the controlled channel attack page fault is that the malicious OS can access and tamper with the page table without restriction, we do not need to make too many changes to the memory layout or access mode. Unlike previous work, in this work, we mainly propose a new defense scheme between MMU traversal and memory.

\begin{table}[htbp!]
  \centering
  \caption{\centering{Performance Comparison of Page Swapping and Non-Swapping Under Attack}}
  \label{table5}
  \small
  \renewcommand{\arraystretch}{1.0}
  \setlength{\tabcolsep}{4pt}
  \begin{tabular}{|c|c|c|c|}
    \hline
     Strategies & \thead{{simTicks}/ \\~{simInsts}} & \thead{simOps /\\hostSeconds}  &  \thead{IPS/\\CPI} \\
    \hline
    \hline
    Non-Swapping & \thead{{$3.34 \times 10^{10}$} /\\~{$4.57 \times 10^9$}} & \thead{$4.59 \times 10^9$ /\\67.93}  &  \thead{$1.37 \times 10^{11}$ /\\7.29} \\
    \hline
    Swapping & \thead{{$1.66 \times 10^{13}$}/ \\~{$2.08 \times 10^{10}$}} & \thead{$3.22 \times 10^{10}$ /\\43039.69}  &  \thead{$1.25 \times 10^9$ /\\797.44} \\
    \hline
  \end{tabular}
\end{table}

\subsubsection{\textbf{Motivation Experiments}}
{To verify the attack characteristics of controlled channel attacks, we implemented two sets of controlled channel attacks on the baseline system. One of the attacks cleared the present bit of the page table entry of the relevant page, and then actually put the page into the LRU list and swapped out the page. Another group of attacks simply clears the present bit of the PTEs of the page in question without actually swapping the pages.}

{Experimental results, as shown in Table \ref{table5}, demonstrate that an attack involving page swapping incurs significantly higher performance overhead compared to an attack that merely clears the present bit. Prior research has established that simply restricting page access by clearing the present bit can effectively execute an attack. Conversely, performing subsequent operations such as deleting PTEs and swapping pages after clearing the present bit introduces substantial and readily detectable overhead. Consequently, the controlled-channel attack, as analyzed, avoids actual page swapping post-present bit clearing, thereby substantially mitigating attack-induced overhead. To facilitate subsequent removal of access restrictions, PTEs are retained in memory. The attack characteristics of controlled-channel attacks are detailed in Section \ref{theorymotivation}.}

\section{controlled-channel attack revisited} \label{controlled-channelAttack}

In this section, we briefly introduce controlled-channel attacks against SGX systems that utilize page faults as a controllable side-channel attack technique. This type of attack enables a malicious OS to infer sensitive computations and data within the TEE. We conduct a focused discussion and analysis of existing SGX side-channel attack defense techniques, and point out the limitations of these techniques.

\subsection{Controlled Side-Channel Attacks}
In the context of Intel SGX, controlled channel attacks exploit page faults as a controllable side channel. A malicious OS can manipulate the page tables of an enclave program, setting the present bits in PTEs and monitoring page faults to determine which memory pages the enclave program intends to access.
As shown in Figure \ref{Fig2}.
Step \textcircled{1}, The adversary sets up an attack trap by maliciously clearing the present bit of a page table entry in the cached page table. Although the present bit is cleared, the content of the page table entry still resides in memory.
Step \textcircled{2}: When the MMU searches for a page table entry and finds that the present bit is 0 (indicating that the searched content is absent), it mistakenly thinks that the required page is not in memory. Therefore, it cannot proceed with step \textcircled{3}.
{Step \textcircled{4}: The MMU requests the OS to handle the page fault, transferring control to the OS and ultimately revealing the page number where the fault occurred. The attacker can modify the corresponding address in the IDT (Interrupt Descriptor Table) to gain access traces of a specific memory page, install a page fault handler, and thus track its control flow or data flow at the page level.

\begin{figure}[!htbp]
\centering
\begin{minipage}{0.45\textwidth} 
    \centering
    \includegraphics[width=\textwidth]{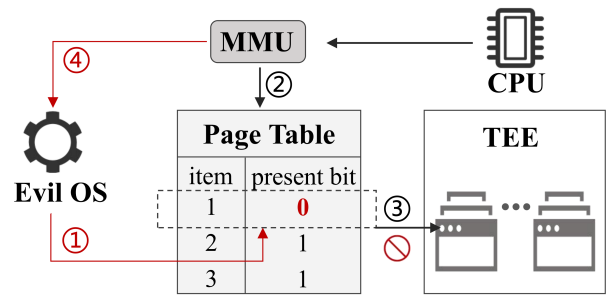} 
    \caption{controlled side-channel attack.} 
    \label{Fig2}
\end{minipage}
\end{figure}

{Since a malicious OS can manipulate the page tables of a protected enclave program, it can determine which memory pages the enclave program intends to access by setting reserved bits in the PTEs and monitoring page faults. Therefore, controlled channel attacks leveraging page faults can be summarized as follows:}

\begin{enumerate}
\item[(1)]{Assume that the adversary attacker can infer the base address of the loaded binary file offline.} 
\item[(2)]{A malicious OS (adversarial attacker) gains unrestricted access to page tables. It can directly modify its present bits to restrict access to specific code or data pages, thereby inducing page faults within the TEE.} 
\item[(3)]{When an application executing in the TEE encounters a page access fault, the shielding system will forward the error request to the OS and report the faulting page address information to the OS.} 
\item[(4)]{The OS then records the relevant information of the faulted page, subsequently restores the page table entry, and removes the restrictions on the page.} 
\item[(5)]{The OS collects page access pattern information to infer privacy-sensitive data flow or control flow information of user applications.} 
\end{enumerate}


Characteristics of Controlled Side-Channel Page Fault Attacks. Analysis of the controlled side-channel page fault attack process reveals five key characteristics:
\begin{enumerate}
    \item[(1)]{Unrestricted Page Table Access. Attackers possess unrestricted access to and modification capabilities of page tables.}
    \item[(2)]{IDT-Based Attack Handler Installation. The attack installs its page fault handler by overwriting corresponding addresses in the Interrupt Descriptor Table (IDT), from which it acquires page access patterns.} 
    \item[(3)]{Persistent PTEs. Even when the present bit in a page table entry is cleared, the page table entry remains stored in memory.} 
    \item[(4)]{Minimization of Attack Overhead. To reduce collateral impact during the attack, such as latency performance degradation, attackers only clear the present bit of PTEs without actual page swapping.} 
    \item[(5)]{Present Bit Restoration. After acquiring page access patterns, the page fault handler restores the present bit, removing access restrictions on the page.} 
\end{enumerate}

\section{Related Work}
\label{Relatedwork}
In terms of detecting the additional execution time of anomalies and interruptions caused by attacks, Chen et al. proposed a software framework, D\'{e}j\`{a} Vu, which implements a novel and reliable software reference clock, which can check the program execution time at the path granularity in its control flow diagram, enabling masking execution to detect such privileged side-channel attacks.

To hide control flows and data flows by reordering the code layout within the Enclave, Shih et al. proposed shih2017t \cite{shih2017t} in 2017, which compiles Enclave applications into a set of TSX transactions. This structure allows page faults to be handled first by the transaction abort handler, which effectively conceals fault addresses from the OS. Apart from shih2017t, another known software-based defense against page-fault side channels involves masking page access patterns from the untrusted OS. This solution was demonstrated on encryption libraries and requires manual annotation, resulting in significant overhead \cite{shinde2016preventing}.

In terms of hiding an application’s memory access patterns through noise injection or ORAM techniques, Zhang et al. proposed Klotski \cite{zhang2020klotski} in 2020, an efficient obfuscated execution technique that achieves a tunable trade-off between security and performance to mitigate controlled-channel attacks. At a high level, Klotski simulates a secure memory subsystem. It employs an enhanced ORAM protocol to load code and data into two software caches of configurable size, which are periodically re-randomized at configurable intervals.More importantly, Klotski incorporates several optimizations to reduce the performance overhead caused by software-based address translation and software cache replacement. InvisiPage \cite{invisipage} introduces a novel hardware design similar to Klotski, where programs inside an enclave can populate their own page tables, with their page access patterns protected by the ORAM protocol. While hardware implementation can offer better performance, Klotski is applicable to existing hardware, and all of Klotski’s compile-time optimizations are also compatible with InvisiPage.

Shinde et al. \cite{shinde2016preventing}. proposed a method called deterministic multiplexing that combines sensitive code and data on a single page to hide page access patterns. However, this method has substantial performance overhead (over 4000$\times$) and requires manual optimization. Other works, such as OBLIVIATE \cite{obliviate}, ZeroTrace \cite{zerotrace}, and OBFUSCURO \cite{obfuscuro}, use ORAM protocols to obfuscate memory access. ZeroTrace only considers data oblivion, while OBLIVIATE provides a file system for Enclaves but does not eliminate controlled-channel execution. OBFUSCURO obfuscates code and data access, offering robust protection against cache and timing side-channel attacks. Still, it only supports small programs (up to 8 KB), making it impractical for real-world applications, with significant performance overhead (55$\times$ on its custom benchmarks).

Zigzagger \cite{inferring} obfuscates a set of branch instructions into a single indirect branch to prevent branch-based side-channel attacks, but it only protects code execution and is defeated by some fine-grained attacks. ENCLANG \cite{compiler} obfuscates leaf functions that do not call other functions. DR. SGX \cite{brasser2019dr} continuously re-randomizes all enclave data at the cache line granularity during enclave execution. However, it only focuses on data access and does not obfuscate code execution.

\section{Design} \label{design}
\subsection{Threat Model}

{We follow the conventional hardware-rooted TEE, i.e., Intel SGX.
Similar to the threat model of previous works \cite{shih2017t, Controlled-Channel-Attacks, shinde2016preventing, oleksenko2018varys}, we first assume that the OS is untrusted. The OS is capable of managing enclave memory pages, although it cannot observe their contents. Whenever an enclave program attempts to access an unmapped page, the OS receives a page fault for processing; the handler then either remaps the page and resumes the program, or generates an access violation error. In other words, all other components in the software stack, except the TEE itself, are considered untrusted. Second, we assume that the adversary can perform a detailed analysis of the target enclave program's source code and/or binaries, and derive memory access patterns. However, we assume that the adversary cannot arbitrarily execute the target enclave program. Fourth, the attack relies solely on a noise-free side channel: page fault information. Other noisy side channels, including cache and memory bus, are beyond the scope of this paper.
}

{Based on the above assumptions, the untrusted OS can manipulate the present bit tag in the page table entry of interest, thereby triggering a page fault interrupt to preempt program execution. By overwriting the corresponding address in the Interrupt Descriptor Table (IDT) to install the attacker's page fault handler, it can track the control flow or data flow at the page level or cache line level.}



\subsection{Defense Tree Based on MT} 

{
Most existing research findings primarily serve a detection function (i.e., detecting whether a page has been tampered with by a malicious OS) rather than a defensive one (i.e., they cannot directly bypass the malicious OS to restore the tampered page or directly access its content).
Therefore, our goal is to design a mechanism that ensures the MMU can still use the physical base address to locate the accessed physical page even if a malicious OS tampers with the present bit of the page table entry. Inspired by the traditional MT architecture, we adopt a similar tree structure to organize the present bits and physical base addresses of PTEs for integrity protection. We refer to this defense tree as the DD-Tree. To enhance lookup efficiency, the defense tree is designed as a completely balanced tree, where each leaf node stores the present bit and physical base address of a page table entry, and non-leaf nodes store the combined hash values of their child nodes, typically computed by concatenating the hash values of its child nodes. Similar to the MT, the defense tree grows in this manner until it converges to the root node. The root node of the defense tree is stored in a TEE, such as Intel SGX, and never leaves it. The root node serves as a robust basis for verifying the integrity of the data stored in the leaf nodes, namely the present bits and physical base addresses.
}


Another key design feature of the defense tree is that we use the linearized virtual address (the starting address of the virtual page) as the index ID of the leaf node. When the MMU performs address translation and needs to verify whether a page table entry has been tampered with, it can efficiently locate the corresponding leaf node in the Defense Tree, which protects the page table entry mapped to the specific virtual address.
The defense mechanism verifies whether its hash matches the hash stored in its parent node and recursively checks the integrity of parent nodes up to the root. During an update, all nodes along the path from the leaf node to the root node are updated accordingly.

\subsection{More Details of The Defense Tree}

{In the DD-Tree, each leaf node is responsible for protecting the integrity of a single page table entry. Due to the large number of PTEs, if a single complete binary tree structure is used for storage, the tree's height becomes very high, which can reduce execution efficiency. Therefore, to improve the execution efficiency of integrity verification, as well as inserting and deleting tree nodes, we limit the height of the DDT-Tree to no more than 8 layers.
To address the issue of insufficient storage capacity in a single tree, we introduce the defense forest. For detailed descriptions, please refer to Section \ref{DefenseForest}.
}


\begin{figure}[!htbp]
\centering
\begin{minipage}{0.45\textwidth} 
    \centering
    \includegraphics[width=\textwidth]{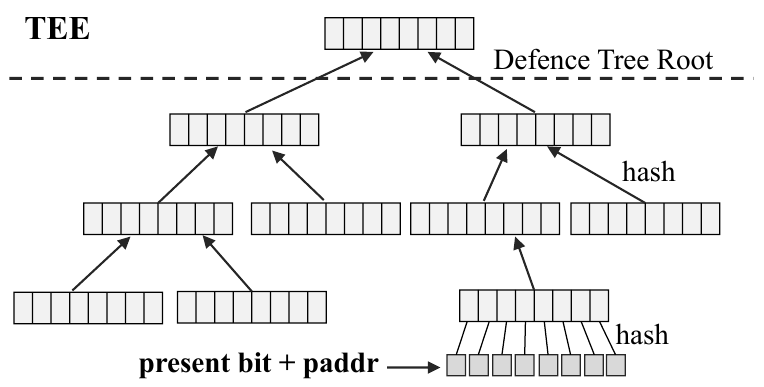} 
    \caption{Defense Tree (Part of the node)} 
    \label{Fig3}
\end{minipage}
\end{figure}

\subsubsection{\textbf{Complete 8-ary Tree}}
{Let $m$ denote the number of hash nodes. We employ a balanced $m$-ary defense tree ($m$-ary-DD-Tree) as illustrated in Figure \ref{Fig3}.  Based on the configured parameters, we can determine that in terms of memory consumption, the indexing overhead is $\frac{1}{m-1}$ times the memory required to protect the data, while the memory required for storing hash values accounts for of the total protected memory. The total additional memory overhead is $\frac{1}{m-1} + \frac{1}{m}$ times the original data size (e.g., 26.8\% when $m=8$), which is considerably lower than the 50\% overhead of a binary tree. Each verification process requires $log_m(N)$ hash verifications, with the cost of each hash computation proportional to $m$(i.e., the number of child nodes being hashed).Analysis reveals that existing systems permit the verification of integrity for a large number of PTEs using only a very small amount of secure storage (typically 128 to 160 bits). Therefore, we consider these cost overheads to be acceptable.}

\subsubsection{\textbf{Defense Forest Using PUD Value as Tree ID}} \label{DefenseForest}
To further optimize the defense mechanism for large-scale data, we organize the defense trees into a forest(DD-Tree-Forest, using the PUD address value as the tree ID. This approach shortens the verification path and improves the efficiency of integrity validation. While increasing the number of defense trees requires additional secure memory to store root hashes, this overhead remains acceptable within the design constraints.

In an OS, the page table is responsible for managing the mapping between virtual addresses and physical addresses. Modern systems commonly use multi-level page tables. For example, in a 32-bit x86 system, the kernel primarily adopts a two-level page table for virtual memory addressing. However, in 64-bit x86 systems, due to the significantly larger virtual address space required by processes, a four-level page table structure is used, consisting of the Page Global Directory (PGD), Page Upper Directory (PUD), Page Middle Directory (PMD), and Page Table (PT). Essentially, a multi-level page table is structured like a multi-way tree, which is why it is often referred to as a page table tree. In multi-level page tables, PTEs inherently exist in a tree-like structure.In the Linux kernel’s memory management mechanism, PGD is the top-level directory that manages page tables. Each application can have one or multiple PGDs, and each process maintains its own independent address space. To ensure independent memory mappings for each process, every process has a separate page table.When protecting the victim bits and physical base addresses of PTEs, a common approach is to store these attributes of entries belonging to the same PGD in the leaf nodes of a defense tree. However, such a defense tree can still be relatively large. To reduce the lookup complexity and improve the efficiency of verification and updates, we further optimize the defense tree structure by grouping PTEs based on the same PUD. In this optimized design, the victim bits and physical base addresses of entries under the same PUD are stored in the leaf nodes of a single defense tree, while the PUD address itself is treated as the defense tree ID, which is stored in the root node. The defense tree ID represents that all protected PTEs belong to the same PUD.

This method effectively decomposes the original defense tree into multiple defense trees corresponding to different PUDs, which we refer to as a defense forest, as illustrated in Figure 4. The roots of the decomposed defense treesmust still be stored in the TEE’s secure memory. Although this defense forest consumes more secure memory, it significantly reduces the size of individual defense trees, thereby improving the efficiency of integrity verification and updates.

In modern OS, $cr3$ register points to the unique PGD of the process, which stores the physical memory address of the PGD. The $cr3$ register serves as the entry point of the paging mechanism, meaning the PGD determines the entire virtual address space layout. In the Linux kernel, the top-level PGD is stored in the $pgd$ variable within the $mm_struct$ of $task_struct$,where $mm_struct$ and $task_struct$ are both structs. When the CPU accesses the virtual memory of a process, it obtains the physical address of the PGD from $cr3$ and performs address translation by sequentially accessing PGD, PUD, PMD, and PT in physical memory, then locating the final physical memory address based on the offset in the virtual address.
In our scheme, when organizing the protection of a new page table entry, we can obtain its PGD address value by accessing the $cr3$ register and then determine the PUD address from the PGD. The PUD address is used to determine which Defense Tree in the Defense Forest should protect this page table entry, and the entry is then added as a leaf node to that tree. If no Defense Tree with the corresponding PUD value exists, a new Defense Tree is created within the Defense Forest. During Defense Tree search, update, and integrity verification, the Tree ID enables precise and efficient location within the Defense Forest.

Fortunately, in the gem5 simulator, we can obtain the PGD and PUD of the page table for the currently accessed virtual address by using the $cr3$ register with the command:
\begin{center}
    $CR3$ $\texttt{cr3 = }$ \\
    $\texttt{tc -> readMiscRegNoEffect(misc\_reg::Cr3)}$
\end{center}
This facilitates the implementation of our Defense Forest design.

\begin{figure}[!htbp]
\centering
\begin{minipage}{0.45\textwidth} 
    \centering
    \includegraphics[width=\textwidth]{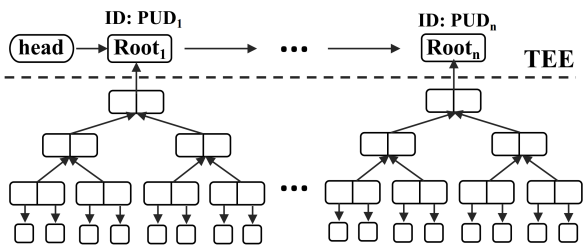} 
    \caption{Defense Forest} 
    \label{Fig4}
\end{minipage}
\end{figure}

\subsection{Creation and Addition of the Defense Tree}

{The primary challenge in constructing a defensive tree (DD-Tree) lies in efficiently creating, adding, and deleting each node. To reduce memory overhead and improve verification efficiency, our implementation does not protect all PTEs. In this work, due to its privileged execution environment, the system PTE has higher stability requirements and a lower attack surface, and we only protect the application PTE, not the system program PTE.}
When a new PTE is created (i.e., when a physical page is swapped from external storage into memory during program execution), our designed system automatically adds the corresponding leaf node to the DD-Tree. Before introducing our scheme, it is essential to understand that when the CPU accesses a physical page that is not present in memory, it attempts memory access twice, as illustrated in Figure \ref{Fig55}. The detailed steps of the CPU accessing a physical page that is not present in memory are as follows:


\begin{enumerate}
\item[(1)]{The MMU first checks the TLB cache, as shown in Figure \ref{Fig55}(\textcircled{1}).}
\item[(2)]{The TLB misses, the Translation Walk Unit (TWU) in the MMU looks up the page table for the mapping, as shown in Figure \ref{Fig55}(\textcircled{2}).}
\item[(3)]{The page table lookup determines that the page is not loaded in memory, a page fault is triggered, as shown in Figure \ref{Fig55}(\textcircled{3}).}
\item[(4)]{The page fault is handled by the OS, which loads the page from external storage into memory and updates the page table and TLB, as shown in Figure \ref{Fig55}(\textcircled{4}).}
\item[(5)]{The processor state is restored, and the interrupted program resumes execution.}
\item[(6)]{The CPU retries accessing the same physical address that initially caused the page fault, as shown in Figure \ref{Fig55}(\textcircled{5}). Since the new PTE and cache have been created and both PT and TLB have been updated, the second memory access successfully hits the TLB cache.}
\end{enumerate}

At this point, the CPU has made two memory access attempts, involving at least two TLB lookups and one page table lookup. These two attempts occur consecutively. During the first memory access attempt, the MMU determines which PTE the OS intends to create. Since this is a new PTE, it is also the PTE that the defense tree is preparing to protect. During the second memory access attempt, the MMU now knows that the PTE has been successfully created, and its complete information can be retrieved from the page table or TLB. The process of two memory access attempts, the MMU knows that the state of the PTE is different, so in order to ensure that the newly loaded PTE is accurately integrated into the DD tree, we introduce two steps: Pre-Addition and Formal Addition.

\subsubsection{\textbf{Pre-Addition}}
During the first TLB and page table lookup, the memory access fails. At this point, the OS begins to load the requested page from external storage into memory and creates the corresponding PTE and cache. However, the MMU only knows that the PTE creation process has started but does not know when it will be completed or the complete PTE information. In Figure \ref{Fig55}, the MMU does not know exactly when step \textcircled{4} finishes.Thus, at this stage, the defense tree prepares to protect the PTE. In our design, when the CPU accesses a physical page that is not present in memory and both the TLB cache lookup and page table traversal fail, the MMU pre-adds the virtual address into the defense tree.

\begin{figure}[!htbp]
\centering
\begin{minipage}{0.45\textwidth} 
    \centering
    \includegraphics[width=\textwidth]{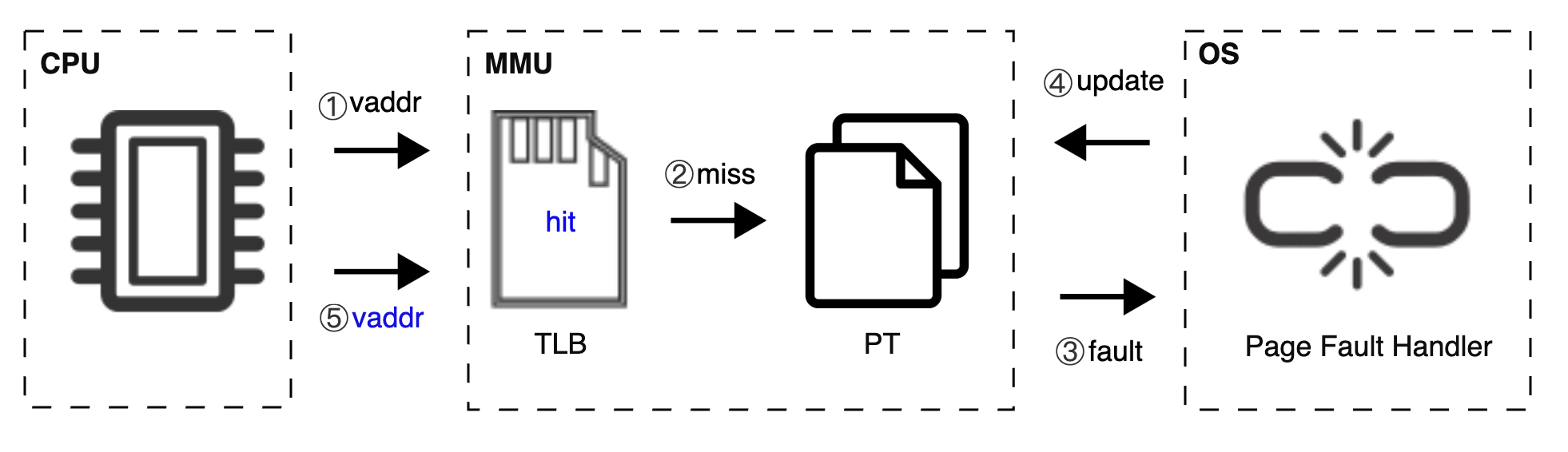} 
    \caption{Addition of the Defense Tree. \textcircled{1}: CPU first memory access attempt; \textcircled{2}: TLB miss; \textcircled{3}: Trigger page fault interrupt; \textcircled{4}: OS updates TLB and PT; \textcircled{5}: CPU second memory access attempt and hit TLB cache. Pre-Addition after \textcircled{3}, Formal Addition after \textcircled{5}} 
    \label{Fig55}
\end{minipage}
\end{figure}

\subsubsection{\textbf{Formal Addition}}
During the first memory access attempt, the MMU fails to locate a mapping in the TLB and page table. The OS handles the page fault, loads the page, assigns a physical address, and updates the TLB and the page table. When the CPU retries accessing the virtual address, the MMU follows Figure \ref{Fig55}(\textcircled{5})and successfully hits in the TLB. This indicates that the PTE has been successfully created and that the MMU can now retrieve its complete information, including the newly assigned physical address. At this moment, the defense scheme formally integrates the PTE into the DD-Tree. A leaf node containing the PTE metadata (that is, victim bits and physical address information) is officially added to the corresponding defense tree ID. If there is no existing ID for the defense tree that matches the PUD value of the PTE, a new defense tree is created within the defense forest.

This two-step process ensures that when the TLB is accessed the second time, the OS has already completed memory allocation, allowing the MMU to obtain the complete page table entry information.
Once a new leaf node is added, all nodes along the path to the root are updated. When the OS swaps out a page to free memory, it updates the corresponding page table entry, and the Defense Tree removes the corresponding leaf node, updating the tree accordingly.

\subsection{The Bridgehead MMU and Its Workflow}\label{Workflow}

{The MMU serves as a bridge between CPU addressing and OS memory management, functioning as a critical 'gateway'. While CPUs execute accesses using physical addresses, modern systems only provide virtual address spaces. The MMU's core function is translating CPU virtual addresses to physical addresses, which relies on page tables rooted at the Page Global Directory (PGD). Although the MMU enforces process isolation and security controls, its page tables are maintained by the OS. This OS-dependence for page table management makes MMU PTEs vulnerable to OS-based side-channel attacks.}

\begin{figure}[!htbp]
\centering
\begin{minipage}{0.45\textwidth} 
    \centering
    \includegraphics[width=\textwidth]{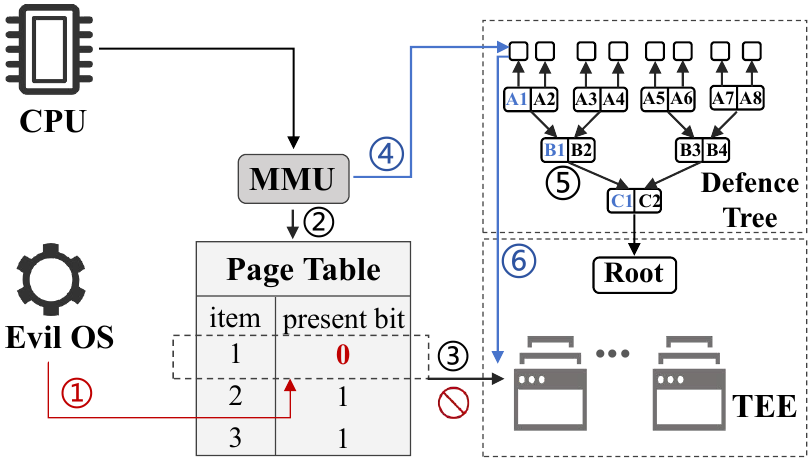} 
    \caption{Detection and Defense Against Controlled Channel Attacks: \textcircled{1}: Attack; \textcircled{2}: MMU Traversing Page Table; \textcircled{3}: Address Translation Failure; \textcircled{4}: Find the corresponding defense tree and corresponding leaf node; \textcircled{5}: Integrity Check; \textcircled{6}: Find the memory page from the physical address in the leaf node} 
    \label{Fig66}
\end{minipage}
\end{figure}

{In this work, we consider the MMU to be a critical component to defend against malicious OS-controlled channel attacks. When the MMU encounters a page table entry (PTE) that exists in the page table but has its present bit set to $0$ (a key characteristic of controlled-channel attacks), it indicates the PTE cannot be fully trusted and may be an attack trap set by the OS. Before requesting OS intervention, the MMU triggers our proposed defense mechanism as illustrated in Figure \ref{Fig66}.}



{In this work, when facing controlled side-channel attacks against PTEs initiated by a malicious OS, the defensive workflow of our 'gateway' MMU is as follows:}

\begin{enumerate}
\item[(1)]{\textbf{Attack:} The malicious OS clears the present bit of a PTE (item 1) in the page table, as shown in Figure \ref{Fig66}(\textcircled{1}).}
\item[(2)]{\textbf{Waiting:} The OS waits for the MMU to request a page fault interrupt for the corresponding PTE.}
\item[(3)]{\textbf{CPU addressing:} The CPU attempts to access the physical memory page, requiring the MMU to translate the virtual address.}
\item[(4)]{\textbf{MMU Traversal:} As shown in Figure \ref{Fig66} (\textcircled{2}), the MMU searches the page table (ignoring the TLB lookup process), finds the matching PTE (item 1), and detects that its present bit is set to 0. Address translation is temporarily halted, as shown in Figure \ref{Fig66} (\textcircled{3}).}
\item[(5)]{\textbf{MMU Self-Check:} Unlike traditional behavior, the MMU does not immediately trigger a page fault interrupt, thus preventing the OS from taking control of the process. Instead, as illustrated in Figure \ref{Fig66} (\textcircled{4}), the MMU accesses the $cr3$ register to locate the PGD and PUD. Using the PUD address, it traverses the defense forest, identifies the corresponding defense tree by comparing tree IDs, linearizes the virtual address, finds the corresponding leaf node, and verifies the integrity of the hash path from the leaf to the root, as shown in Figure \ref{Fig66} (\textcircled{5}).}
\item[(6)]{\textbf{Integrity Check Failure:} If the integrity verification fails, it confirms that the “PTE exists in the page table but has a present bit of 0” scenario is due to an attack. The MMU detected the occurrence of a controlled-channel attack.}
\item[(7)]{\textbf{Bypassing OS in Page Fault Handling:} If the MMU were to trigger a page fault interrupt to the OS, it would likely fall into the OS’s trap. Instead, after detecting an attack, the MMU retrieves the stored physical address from the leaf node of the defense tree, uses this address to locate the corresponding physical page within the TEE (e.g., SGX Enclave), and sets the present bit of PTE (item 1) to 1, as shown in Figure \ref{Fig66} (\textcircled{6}).}
\item[(8)]{\textbf{Attack Failure:} The MMU successfully avoids the OS’s trap. Since the OS never receives a page fault interrupt for the attacked PTE, it cannot infer the application’s memory access patterns.}
\end{enumerate}

\textbf{Case of Integrity Verification Success.} 
{Under non-attack execution scenarios (e.g., when the OS exhausts physical memory, requires page reclamation, or experiences page loading delays), page tables may contain PTEs with their present bits cleared (0). If the MMU's defensive tree verification detects no anomalies, this indicates the cleared present bit results from legitimate memory operations rather than side-channel attacks. The system should then report this condition to the OS for page table updates before resuming normal execution. Since this doesn't constitute an OS trap, malicious OS cannot infer application memory access patterns through such page faults.
When integrity verification fails, since the essential nature of a page fault is a numerical error in the present bit, the MMU alone cannot resolve the page fault.}

\textbf{Case of Integrity Verification Failure.}
When integrity verification fails, the hash of the leaf node does not match the hash stored in its parent node. This indicates that the abnormal condition - where a page table entry exists in the page table but has its present bit set to 0 - results from unauthorized OS tampering.As previously described in the defense mechanism of the “bridgehead” MMU, the MMU does not rely on the OS to handle this page fault. Instead, it retrieves the stored physical address from the leaf node of the defense tree and uses this address to locate the corresponding physical page within a TEE (e.g., SGX Enclave) to satisfy the CPU’s memory access request. Subsequently, our defense mechanism disrupts the OS attack trap by setting the present bit of the compromised page table entry (item 1) to 1, neutralizing the attack.

The above introduces two cases that lead to the success and failure of integrity verification, as well as different handling methods. When integrity verification fails, the MMU alone can solve the page fault problem because the essence of the page fault error is that the present bit has occurred a numerical error. When integrity verification is successful, it indicates that the page fault error is real and requires the strong memory management capability of the OS to solve it.

\subsection{Security and Efficiency Analysis}
\textbf{Why is it secure?}
The defense tree in our design is essentially a variant of the MT. Collision resistance is one of the core security properties of cryptographic hash functions, ensuring that an attacker cannot feasibly find two different input values that produce the same hash value. The collision resistance of the defense tree is defined as follows: Given a hash function $H$, finding two distinct inputs $x\neq y$, which makes $H(x)= H(y)$ computationally unfeasible.

Assume that a naive defense tree is a binary tree with leaf nodes $n=2^{\mathrm{k}}$ and the height of the tree is $k+1$. Any modification to a leaf node propagates upward, ultimately altering the root hash. If an attacker aims to keep the root hash unchanged, they must satisfy the condition:
\begin{equation}
    H (original\ data\ path) = H (tampered\ data\ path)
\end{equation}

Since the hash value of each layer depends on the hash value of the child node, the attacker needs to find the hash collision at each layer, and its success probability is as follows:
\begin{equation}
    \text{Prob. of Success}\leq\prod_{i=1}^k\text{Layer }i\text{ hash coll. prob.}
\end{equation}

If an attacker attempts to modify a specific leaf node (e.g. changing data block $D_i$ to $D_i^{\prime}$ ), they must compute a new hash value $H(D_i^{\prime})$ and generate new intermediate hash values along the path up to the root, ultimately producing the same original root hash. Due to the collision resistance of the hash function, the probability of successfully forging such a collision chain is negligible.

By verifying the root hash (which is stored in a tamper-proof secure memory and contains information about all other nodes), any modification to a node in the defense tree can be detected, ensuring data integrity.

\textbf{Why is it efficient?}
In the design, the linearized virtual address is used as the ID of the defense leaf node, and the integrity of the data can be verified with a small amount of data (hash path). In order to further improve the verification efficiency of the defense tree, a defense forest with the PUD as the root ID is also designed, and each tree is a multi-fork balanced tree, which can reduce the height of the tree and the length of the hash path.

\section{Evaluation} \label{evaluation}

\subsection{Evaluation  Methodology}
This section describes in detail the experimental methods for evaluating the performance of detecting attacks and defenses.

\subsubsection{\textbf{Simulation Framework}}
{We employ the academically and industrially recognized Gem5 simulator \cite{lowe2021gem5} (a modular, discrete-event-driven computer system simulation platform) for comprehensive performance evaluations. Utilizing the Gem5 V23 architecture, we constructed an x86 full-system simulation to assess the detection and defense methodology proposed in this work.
In our experimental setup within the Gem5 execution environment, the KVM CPU model leverages the host machine's hardware virtualization capabilities to accelerate OS bootstrapping. We utilize Gem5's CPU model switching mechanism to transition to the Timing CPU model. Since other CPU models in Gem5 lack modular software implementation of the MMU, our experiments primarily focus on full-system software simulation with the Timing CPU model - including both MMU and TLB components. To implement our design, we modified the Timing CPU's MMU by altering its page table walk logic.
}

{Under the Time-CPU model, when the MMU module in gem5 executes the $startWalk$ method, it invokes the $sendPackets$ method to send requests to the page table walker and initiate the page table traversal process. During the traversal, the $stepWalk$ method is called to execute page table traversal operations step-by-step. These operations typically involve sending requests to the page table, receiving responses, and performing corresponding actions based on the responses until the entire page table traversal is completed. }

To facilitate the implementation of this design, particularly in a software MMU, we slightly adjusted the logic of the page table traversal of MMU. During the MMU's operation of traversing the page table, the method $stepWalk$ checks the status of the PTE. When $pte.p$ is false, the MMU does not simply hand over control to the OS for handling before proceeding with address translation. Instead, following the detection and defense methodology described in Section \ref{Workflow}, the MMU first verifies the integrity of the defense tree to determine whether the PTE has been potentially attacked. If an attack is detected, the MMU autonomously corrects the erroneous present bit of the PTE without relying on the OS.


In this work, QEMU and PACKER are used to build disk images that provide the OS environment, including file systems, user-space programs, etc.The target of defense and attack will be placed here, and similarly, the SPEC CPU2017\cite{spec2017cpu} image will be added to the file system via PACKER, and the SPEC CPU2017 benchmarks will be run by scripting, mounting the spec cpu2017-1.0.5.iso, configuring the environment and installation.


{
To compile a Linux 5.4.49 kernel that can be tested in the Gem5 simulator. This kernel is capable of executing controlled side-channel attack simulation verification, acting as a malicious OS (attacker). Since page tables are managed by the OS, modifying PTEs of interest in the custom kernel is straightforward. In the absence of defensive measures, the malicious OS overwrites the corresponding addresses in the Interrupt Descriptor Table (IDT) and installs the attacker's page fault handler. Consequently, page miss errors caused by the attacker's modification of PTEs are sent to the OS within the MMU for processing, where page access traces can be captured \ref{table3}. When the system operates under the defense method proposed in this paper, the attacker's page fault handler fails to receive page miss error handling requests for the attacked pages, indicating successful defense. Therefore, in this experiment, the effectiveness of the proposed defense method is evaluated by measuring the success rate of the defense.}

{In Gem5, we simulate an SGX Enclave by partitioning a memory region as a TEE. This memory region is protected against controlled-channel attacks through confidentiality and integrity safeguards. The application's physical memory blocks are stored within this enclave, and to ensure integrity verification of the defensive tree (DD-Tree), its root node is also securely stored in this protected area.}


{Based on the above configurations, our experimental environment has been fully set up. Within this environment, the MMU, file system, and OS kernel have all been customized according to the theoretical analysis and design presented in this paper, forming a computer system capable of meeting the evaluation requirements. The baseline we adopted is a computer system without any protective measures, built using the Gem5 simulator. Specific system parameter settings are shown in Table \ref{table1}.}


\begin{table}[!htbp]
  \centering
  \caption{Baseline System Configuration.}
  \label{table1}
  \begin{tabular}{|l|l|}
    \hline
    \thead{\textbf{Parameter}} & \thead{\textbf{Value}}\\
    \hline
    \hline
    \thead{CPU/Cores/OS/clk\_freq}           & \thead{x86\_64/2/Ubuntu 18.04.2 LTS/3GHz} \\    
    \hline       
    \thead{L1d\_size/L2\_size}          & \thead{32kB/256kB}\\
    \hline
    \thead{Memory size/GCC version}        & \thead{2.923 GB (DDR4)/7.5.0}\\   
    \hline    
  \end{tabular} 
\end{table}

\subsubsection{\textbf{Benchmark}}
{To evaluate the performance impact of our proposed defense technology on the system, we use the benchmark SPEC CPU2017.
To better replicate the attack process, similar to \cite{liu2025hm, zhao2013kiln,chen2020cachetree} we develop a set of benchmarks, as shown in Table \ref{table2}. This set of benchmarks performs search, insertion, and deletion operations on the data structures used in databases and file systems, including search trees, hash tables, sparse graphs, and arrays.
}Additionally, it includes benchmarks capable of accessing at least $n$ virtual addresses ($n$-times). We use this benchmark suite as the target application to conduct controlled-channel attacks and apply the proposed defense strategy to assess both the effectiveness and performance of the defense mechanism.

\begin{figure*}[!htbp]
\centering
\includegraphics[width=0.9\textwidth]{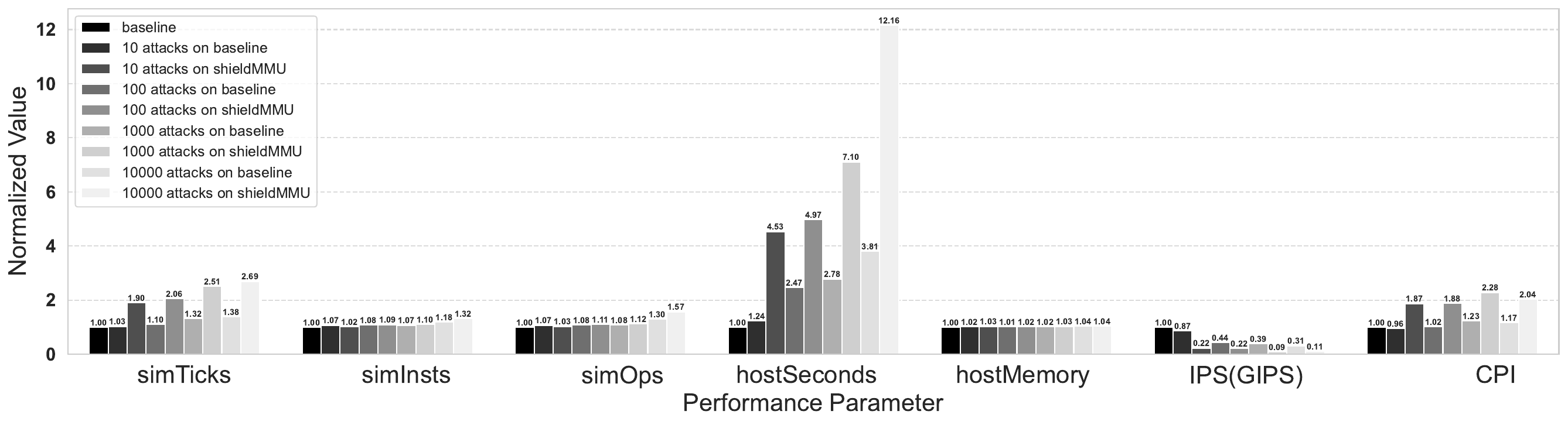} 
\caption{Prformance Metrics Normalized to Baseline. simTicks: Number of ticks simulated (Tick). simInsts: Number of instructions simulated (Count). simOps: Number of ops (including micro ops) simulated (Count). hostSeconds: Real time elapsed on the host (Second). hostMemory: Number of bytes of host memory used (Byte). IPS: Instructions Per Second. CPI: Cycles Per Instruction. Global frequency set at $10^{12}$ ticks per second.} 
\label{Fig9}
\end{figure*}

\subsection{Evaluation Result}
The results of the experiments will be presented in this section.
{We conducted comprehensive experimental validation from multiple perspectives, including side-channel attack resistance, latency performance, and the rationale for selecting the m-ary tree structure.}

\subsubsection{\textbf{Performance Evaluation}}
As shown in the Figure \ref{Fig9}, we conducted $n$-times ($n$ = 10, 100, 1000, 10,000) controlled-channel attacks on an application in both the baseline system and the system with the proposed shieldMMU mechanism.
{In this experiment, we evaluate and compare three configurations: (1) no-attack no-defense (baseline), (2) attack without defense, and (3) attack with defense(shieldMMU).
Based on the aforementioned cases, we conduct a detailed comparative analysis of performance, memory consumption, and efficiency. As predicted by the theoretical analysis in this paper, the application of the proposed defense mechanism within the baseline system does not introduce significant performance overhead.
}

\begin{table}[!htbp]
  \centering
  \caption{Benchmarks Used in Our Experiments.}
  \label{table2}
  \begin{tabular}{|l|l|}
    \hline
    \thead{\textbf{Benchmarks}}     & \thead{\textbf{Description}} \\
    \hline
    \hline
    \thead{$n$-times}  & \thead{Visit the virtual address at least $n$ times.} \\
    \hline
    \thead{BTree}      & \thead{Inserts/deletes nodes in a B-tree.} \\
    \hline
    \thead{Hash}       & \thead{Inserts/deletes entries in a hash table.}\\
    \hline
    \thead{RBTree}     & \thead{Inserts/deletes nodes in a red-black tree.}\\
    \hline
    \thead{SDG }       & \thead{Inserts/deletes edges in a scalable large graph.}\\
    \hline
    \thead{SPS}        & \thead{Random swaps between entries in an array.}\\
    \hline
    \thead{SSCA2}      & \thead{A scalable large graph analysis benchmark.}\\
    \hline
  \end{tabular}
\end{table}



\textbf{Performance.} 
{Under both attack and defense conditions, experimental results show that the clock frequency remains unaffected by either the attack process or the defense mechanism, consistently maintaining a stable rate of $10^{12}$ ticks per second.
As shown in Figure \ref{Fig9}, the number of ticks executed during the attack under undefended conditions increases by $14\%$ compared to the baseline, which executed $3.34 \times 10^{10}$ ticks.
The number of ticks executed under an attack with defense versus an attack without defense doubles. While our proposed shieldMMU mechanism introduces a certain execution overhead, it remains acceptable for security-critical applications. Furthermore, the number of instructions executed remains largely constant across both cases.
Although the shieldMMU mechanism introduces additional runtime overhead, it demonstrates strong defensive capabilities, achieving a protection rate of over $90\%$.
}


\textbf{Efficiency.} 
{
We compute Instructions Per Second (IPS) and Cycles Per Instruction (CPI) using the formulas $IPS = \frac{simInsts \times 10^{12}}{simTicks}$ and $CPI = \frac{simTicks}{simInsts}$ , respectively. 
As shown in Figure \ref{Fig9}, the IPS values decrease by $7\%$ for the attack without defense and by $46\%$ for the attack with defense, indicating a reduction in the number of instructions completed per unit time and a decrease in efficiency for both scenarios. Notably, the impact of the attack on execution efficiency is significantly greater than that of the defense. From a CPI perspective, compared to the baseline, the CPI value increases by $8\%$ for the attack without defense, while the CPI value for the attack with defense is 1.9 times the baseline. Considering the requirements of high-security applications, the performance overhead introduced by the defense mechanism is acceptable.
}

\textbf{Memory.} 
{
The memory overhead increase is maintained below $5\%$ in both attack scenarios (with and without protection) relative to the baseline ( as illustrated in Figure \ref{Fig9}). This marginal increment is deemed acceptable for high-security deployments.
}

\begin{table}[h!]
  \centering
  \caption{SPEC CPU2017 Runtime Comparison.}
  \label{table4}
  \begin{tabular}{|c|c|c|}
    \hline
    \thead{\textbf{benchmark}} & \thead{\textbf{baseline}}   & \thead{\textbf{shieldMMU}}  \\ 
    \hline
    \hline
    \thead{{500.perlbench\_r}/{505.mcf\_r}/ \\~{548.exchange2\_r}/510.parest\_r} & \thead{101.000/107.000 \\~/99.000/73.900}   &  \thead{102.000/107.000/ \\~99.100/74.000} \\
    \hline
   \thead{502.gcc\_r/523.xalancbmk\_r/}           & \thead{0.012/0.562} & \thead{0.036/0.560} \\
    \hline
   \thead{503.bwaves\_r/525.x264\_r/ \\~ 527.cam4\_r}    & \thead{350.000/572.000/ \\~ 285.000}& \thead{350.000/572.000/ \\~ 286.000} \\
        
    \hline
    \thead{519.lbm\_r/541.leela\_r/ \\~ 549.fotonik3d\_r/ 531.deepsjeng\_r/ \\~557.xz\_r}        & \thead{30.400/31.300/\\~ 45.500/ 49.300/ \\~ 67.900} & \thead{30.800/31.400/\\~ 45.500/ 49.400/ \\~ 68.000}  \\    
    \hline  
    
    \thead{602.gcc\_s/631.deepsjeng\_s }       & \thead{0.036/0.005} & \thead{0.036/0.005} \\
    \hline
    \thead{605.mcf\_s/648.exchange2\_s}       & \thead{107.000/99.100} & \thead{109.000/99.100} \\
    \hline
    \thead{619.lbm\_s/ 620.omnetpp\_s/ \\~649.fotonik3d\_s}  & \thead{10.200/21.400/\\~ 42.400}  & \thead{10.900/21.400/\\~ 42.900} \\    
    \hline
  \end{tabular}
\end{table}

\subsubsection{\textbf{Evaluation Results on SPEC CPU2017 Benchmark Suite}}

We evaluate the performance of our proposed ShieldMMU system using the SPEC CPU2017 \cite{bucek2018spec} benchmark suite.
As shown in Table~\ref{table4}, under both the baseline and ShieldMMU configurations, the execution time of nearly all benchmark programs remains unchanged or increases by less than $1\%$. Therefore, the proposed ShieldMMU system provides high security assurance for non-target programs while incurring minimal performance loss, with the overall system performance overhead remaining within an acceptable range.


\begin{figure}[!htbp]
\centering
\begin{minipage}{0.45\textwidth} 
    \centering
    \includegraphics[width=\textwidth]{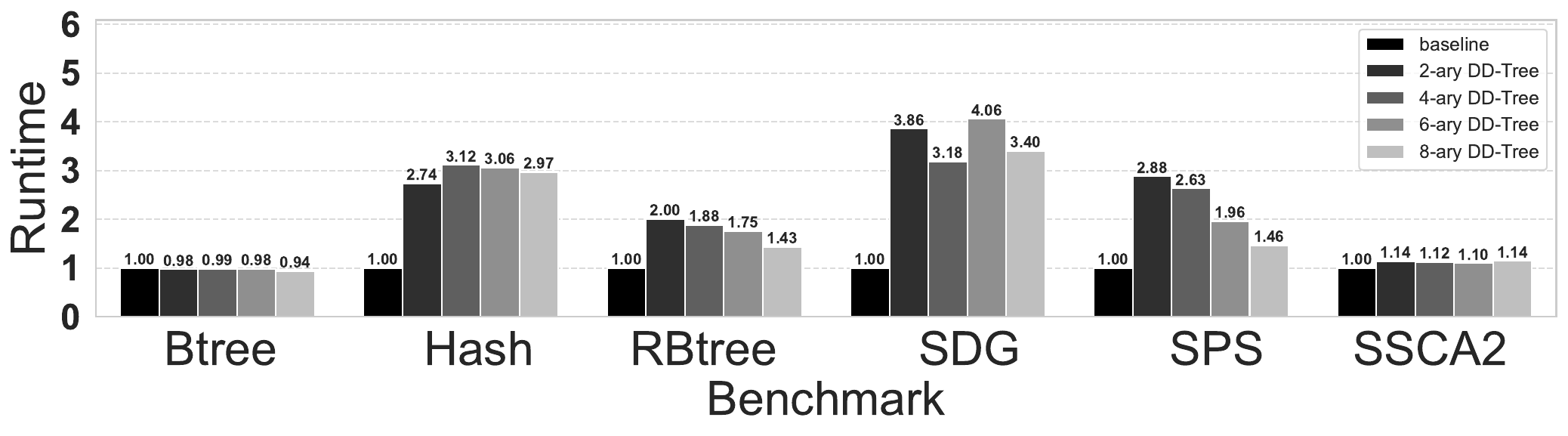} 
    \caption{m-ary DD-Tree Runtime Comparison.} 
    \label{Fig7}
\end{minipage}
\end{figure}

\subsubsection{\textbf{Evaluation Results of m-ary DD-Tree}}
{As illustrated in Figures \ref{Fig7} and \ref{Fig8}, a comparative analysis of runtime and memory overhead among 2-ary, 4-ary, 6-ary, and 8-ary DD trees reveals that the 8-ary DD tree exhibits shorter runtime and reduced memory footprint. While the pointer overhead per non-leaf node in the 8-ary DD tree is higher than in other trees, with a fixed number of leaf nodes, the 8-ary DD tree possesses fewer non-leaf nodes and a lower tree height, thereby achieving more efficient verification and update operations. Consequently, we consider that selecting the 8-ary DD tree as the defense tree for the ShieldMMU mechanism is more efficient and justified.}

\begin{figure}[!htbp]
\centering
\begin{minipage}{0.45\textwidth} 
    \centering
    \includegraphics[width=\textwidth]{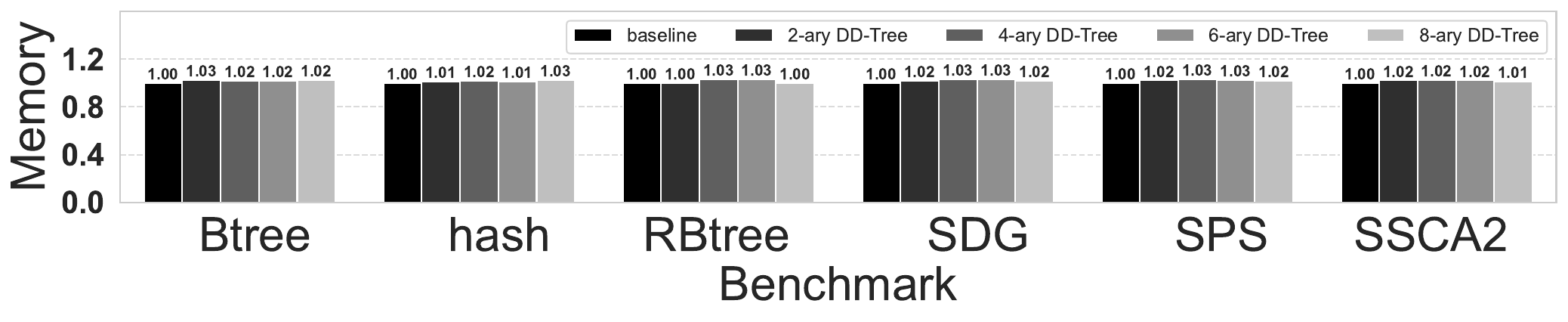} 
    \caption{m-ary DD-Tree Memory Comparison.} 
    \label{Fig8}
\end{minipage}
\end{figure}


\subsubsection{\textbf{Defensive Effect}}
In the table \ref{table3}, \textit{Vaddr Access} represents the number of virtual addresses accessed by the benchmark. \textit{Attack (Leakages)} indicates the number of controlled-channel attack attempts and the number of virtual addresses accessed by the benchmark that were successfully leaked. \textit{shieldMMU Failures (Leakages)} represents the number of virtual addresses accessed by the benchmark that the attacker could still steal after enabling the proposed shieldMMU mechanism. \textit{$100$ visits} and \textit{$1000$ visits} respectively indicate programs that have performed more than 100 and 1000 virtual address accesses. From the final \textit{Success Rate}, the shieldMMU mechanism effectively blocks information leakage in small-scale attacks and demonstrates outstanding defensive performance across benchmarks such as \textit{BTree}, \textit{Hash}, and \textit{SDG}. As the access scale increases, the amount of information leaked after the defense shows a slight increase; however, the defense success rate remains high.

\begin{table}[h!]
  \centering
  \caption{Defensive effect on our benchmarks.}
  \label{table3}
  \small  
  \renewcommand{\arraystretch}{1.0}  
  \begin{tabular}{|c|c|c|c|c|}
    \hline
    \textbf{\thead{Benchmark}} 
    & \shortstack{\thead{\textbf{V-addr}\\\textbf{Count}}} 
    & \shortstack{\thead{\textbf{Attack}\\\textbf{(Leakages)}}} 
    & \shortstack{\thead{\textbf{shieldMMU Failures}\\\textbf{(Leakages)}}}  
    & \shortstack{\thead{\textbf{Success}\\\textbf{Rate}}} \\
    \hline
    \hline
    \thead{{100 accesses}/ \\ {1000 accesses}} & \thead{100/ \\1000}  &  \thead{98/498}  &  \thead{1/7}  &  \thead{99.0\%/98.6\%} \\
    \hline
    \thead{{BTree}/{Hash}} & \thead{49/49}  &  \thead{12/16}  &  \thead{0/0}  &  \thead{100.0\%/100.0\%} \\
    \hline
    \thead{{RBTree}/{SDG}} & \thead{40/45}  &  \thead{17/22}  &  \thead{1/0}  &  \thead{94.1\%/100.0\%} \\
    \hline
    \thead{{SPS}/{SSCA2}} & \thead{50/59}  &  \thead{19/35}  &  \thead{0/0}  &  \thead{100.0\%/100.0\%} \\
    \hline
  \end{tabular}
\end{table}

\section{Conclusion}
To mitigate the threat of side-channel attacks, such as controlled-channel attacks originating from the operating system in isolation systems, we propose a defense mechanism named ShieldMMU. ShieldMMU leverages a DD-Tree structure to protect the integrity of the present bit and physical address in PTEs, enabling the detection of MMU page table lookup events. This allows timely identification of PTEs targeted by controlled-channel attacks. Although the mechanism introduces some performance overhead, it shows strong defensive effectiveness, achieving a detection rate of more than $90\%$.


\bibliographystyle{ACM-Reference-Format}
\bibliography{sample-base}


\begin{thebibliography}{47}


\ifx \showCODEN    \undefined \def \showCODEN     #1{\unskip}     \fi
\ifx \showDOI      \undefined \def \showDOI       #1{#1}\fi
\ifx \showISBNx    \undefined \def \showISBNx     #1{\unskip}     \fi
\ifx \showISBNxiii \undefined \def \showISBNxiii  #1{\unskip}     \fi
\ifx \showISSN     \undefined \def \showISSN      #1{\unskip}     \fi
\ifx \showLCCN     \undefined \def \showLCCN      #1{\unskip}     \fi
\ifx \shownote     \undefined \def \shownote      #1{#1}          \fi
\ifx \showarticletitle \undefined \def \showarticletitle #1{#1}   \fi
\ifx \showURL      \undefined \def \showURL       {\relax}        \fi
\providecommand\bibfield[2]{#2}
\providecommand\bibinfo[2]{#2}
\providecommand\natexlab[1]{#1}
\providecommand\showeprint[2][]{arXiv:#2}

\bibitem[Aga and Narayanasamy(2019)]%
        {invisipage}
\bibfield{author}{\bibinfo{person}{Shaizeen Aga} {and} \bibinfo{person}{Satish
  Narayanasamy}.} \bibinfo{year}{2019}\natexlab{}.
\newblock \showarticletitle{InvisiPage: Oblivious demand paging for secure
  enclaves}. In \bibinfo{booktitle}{\emph{Proceedings of the 46th International
  Symposium on Computer Architecture}}. \bibinfo{pages}{372--384}.
\newblock


\bibitem[Ahmad et~al\mbox{.}(2019)]%
        {obfuscuro}
\bibfield{author}{\bibinfo{person}{Adil Ahmad}, \bibinfo{person}{Byunggill
  Joe}, \bibinfo{person}{Yuan Xiao}, \bibinfo{person}{Yinqian Zhang},
  \bibinfo{person}{Insik Shin}, {and} \bibinfo{person}{Byoungyoung Lee}.}
  \bibinfo{year}{2019}\natexlab{}.
\newblock \showarticletitle{OBFUSCURO: A commodity obfuscation engine on Intel
  SGX}. In \bibinfo{booktitle}{\emph{Network and Distributed System Security
  Symposium}}.
\newblock


\bibitem[Ahmad et~al\mbox{.}(2018)]%
        {obliviate}
\bibfield{author}{\bibinfo{person}{Adil Ahmad}, \bibinfo{person}{Kyungtae Kim},
  \bibinfo{person}{Muhammad~Ihsanulhaq Sarfaraz}, {and}
  \bibinfo{person}{Byoungyoung Lee}.} \bibinfo{year}{2018}\natexlab{}.
\newblock \showarticletitle{OBLIVIATE: A Data Oblivious Filesystem for Intel
  SGX.}. In \bibinfo{booktitle}{\emph{NDSS}}.
\newblock


\bibitem[Anati et~al\mbox{.}(2013)]%
        {anati2013innovative}
\bibfield{author}{\bibinfo{person}{Ittai Anati}, \bibinfo{person}{Shay Gueron},
  \bibinfo{person}{Simon Johnson}, {and} \bibinfo{person}{Vincent Scarlata}.}
  \bibinfo{year}{2013}\natexlab{}.
\newblock \showarticletitle{Innovative technology for CPU based attestation and
  sealing}. In \bibinfo{booktitle}{\emph{Proceedings of the 2nd international
  workshop on hardware and architectural support for security and privacy}},
  Vol.~\bibinfo{volume}{13}. ACM New York, NY, USA.
\newblock


\bibitem[Baumann et~al\mbox{.}(2015)]%
        {baumann2015shielding}
\bibfield{author}{\bibinfo{person}{Andrew Baumann}, \bibinfo{person}{Marcus
  Peinado}, {and} \bibinfo{person}{Galen Hunt}.}
  \bibinfo{year}{2015}\natexlab{}.
\newblock \showarticletitle{Shielding applications from an untrusted cloud with
  haven}.
\newblock \bibinfo{journal}{\emph{ACM Transactions on Computer Systems (TOCS)}}
  \bibinfo{volume}{33}, \bibinfo{number}{3} (\bibinfo{year}{2015}),
  \bibinfo{pages}{1--26}.
\newblock


\bibitem[Berbecaru and Albertalli(2008)]%
        {Berbecaru_Albertalli_2008}
\bibfield{author}{\bibinfo{person}{Diana Berbecaru} {and} \bibinfo{person}{Luca
  Albertalli}.} \bibinfo{year}{2008}\natexlab{}.
\newblock \showarticletitle{On the performance and use of a space-efficient
  merkle tree traversal algorithm in real-time applications for wireless and
  sensor networks}. In \bibinfo{booktitle}{\emph{2008 IEEE International
  Conference on Wireless and Mobile Computing, Networking and Communications}}.
  IEEE, \bibinfo{pages}{234--240}.
\newblock


\bibitem[Brasser et~al\mbox{.}(2019)]%
        {brasser2019dr}
\bibfield{author}{\bibinfo{person}{Ferdinand Brasser}, \bibinfo{person}{Srdjan
  Capkun}, \bibinfo{person}{Alexandra Dmitrienko}, \bibinfo{person}{Tommaso
  Frassetto}, \bibinfo{person}{Kari Kostiainen}, {and}
  \bibinfo{person}{Ahmad-Reza Sadeghi}.} \bibinfo{year}{2019}\natexlab{}.
\newblock \showarticletitle{DR. SGX: Automated and adjustable side-channel
  protection for SGX using data location randomization}. In
  \bibinfo{booktitle}{\emph{Proceedings of the 35th Annual Computer Security
  Applications Conference}}. \bibinfo{pages}{788--800}.
\newblock


\bibitem[Bucek et~al\mbox{.}(2018)]%
        {bucek2018spec}
\bibfield{author}{\bibinfo{person}{James Bucek}, \bibinfo{person}{Klaus-Dieter
  Lange}, {and} \bibinfo{person}{J{\'o}akim v. Kistowski}.}
  \bibinfo{year}{2018}\natexlab{}.
\newblock \showarticletitle{SPEC CPU2017: Next-generation compute benchmark}.
  In \bibinfo{booktitle}{\emph{Companion of the 2018 ACM/SPEC International
  Conference on Performance Engineering}}. \bibinfo{pages}{41--42}.
\newblock


\bibitem[Checkoway and Shacham(2013)]%
        {Iago}
\bibfield{author}{\bibinfo{person}{Stephen Checkoway} {and}
  \bibinfo{person}{Hovav Shacham}.} \bibinfo{year}{2013}\natexlab{}.
\newblock \showarticletitle{Iago attacks: why the system call API is a bad
  untrusted RPC interface}.
\newblock \bibinfo{journal}{\emph{SIGARCH Comput. Archit. News}}
  \bibinfo{volume}{41}, \bibinfo{number}{1} (\bibinfo{date}{March}
  \bibinfo{year}{2013}), \bibinfo{pages}{253–264}.
\newblock
\showISSN{0163-5964}
\urldef\tempurl%
\url{https://doi.org/10.1145/2490301.2451145}
\showDOI{\tempurl}


\bibitem[Chen et~al\mbox{.}(2017)]%
        {chen2017detecting}
\bibfield{author}{\bibinfo{person}{Sanchuan Chen}, \bibinfo{person}{Xiaokuan
  Zhang}, \bibinfo{person}{Michael~K Reiter}, {and} \bibinfo{person}{Yinqian
  Zhang}.} \bibinfo{year}{2017}\natexlab{}.
\newblock \showarticletitle{Detecting privileged side-channel attacks in
  shielded execution with D{\'e}j{\'a} Vu}. In
  \bibinfo{booktitle}{\emph{Proceedings of the 2017 ACM on Asia Conference on
  Computer and Communications Security}}. \bibinfo{pages}{7--18}.
\newblock


\bibitem[Chen et~al\mbox{.}(2020)]%
        {chen2020cachetree}
\bibfield{author}{\bibinfo{person}{Zhengguo Chen}, \bibinfo{person}{Youtao
  Zhang}, {and} \bibinfo{person}{Nong Xiao}.} \bibinfo{year}{2020}\natexlab{}.
\newblock \showarticletitle{Cachetree: Reducing integrity verification overhead
  of secure nonvolatile memories}.
\newblock \bibinfo{journal}{\emph{IEEE Transactions on Computer-Aided Design of
  Integrated Circuits and Systems}} \bibinfo{volume}{40}, \bibinfo{number}{7}
  (\bibinfo{year}{2020}), \bibinfo{pages}{1340--1353}.
\newblock


\bibitem[Cheng et~al\mbox{.}(2013)]%
        {cheng2013appshield}
\bibfield{author}{\bibinfo{person}{Yueqiang Cheng}, \bibinfo{person}{Xuhua
  Ding}, {and} \bibinfo{person}{Robert Deng}.} \bibinfo{year}{2013}\natexlab{}.
\newblock \showarticletitle{Appshield: Protecting applications against
  untrusted operating system}.
\newblock \bibinfo{journal}{\emph{Singaport Management University Technical
  Report, SMU-SIS-13}}  \bibinfo{volume}{101} (\bibinfo{year}{2013}).
\newblock


\bibitem[Criswell et~al\mbox{.}(2014)]%
        {criswell2014virtual}
\bibfield{author}{\bibinfo{person}{John Criswell}, \bibinfo{person}{Nathan
  Dautenhahn}, {and} \bibinfo{person}{Vikram Adve}.}
  \bibinfo{year}{2014}\natexlab{}.
\newblock \showarticletitle{Virtual ghost: Protecting applications from hostile
  operating systems}.
\newblock \bibinfo{journal}{\emph{ACM SIGARCH Computer Architecture News}}
  \bibinfo{volume}{42}, \bibinfo{number}{1} (\bibinfo{year}{2014}),
  \bibinfo{pages}{81--96}.
\newblock


\bibitem[Elbaz et~al\mbox{.}(2009)]%
        {Elbaz_Champagne_Gebotys_Lee_Potlapally_Torres_2009}
\bibfield{author}{\bibinfo{person}{Reouven Elbaz}, \bibinfo{person}{David
  Champagne}, \bibinfo{person}{Catherine Gebotys}, \bibinfo{person}{Ruby~B.
  Lee}, \bibinfo{person}{Nachiketh Potlapally}, {and} \bibinfo{person}{Lionel
  Torres}.} \bibinfo{year}{2009}\natexlab{}.
\newblock \bibinfo{booktitle}{\emph{Hardware Mechanisms for Memory
  Authentication: A Survey of Existing Techniques and Engines}}.
\newblock \bibinfo{pages}{1–22}.
\newblock
\urldef\tempurl%
\url{https://doi.org/10.1007/978-3-642-01004-0_1}
\showDOI{\tempurl}


\bibitem[Freij et~al\mbox{.}(2021)]%
        {Freij_Zhou_Solihin_2021}
\bibfield{author}{\bibinfo{person}{Alexander Freij}, \bibinfo{person}{Huiyang
  Zhou}, {and} \bibinfo{person}{Yan Solihin}.} \bibinfo{year}{2021}\natexlab{}.
\newblock \showarticletitle{Bonsai merkle forests: Efficiently achieving crash
  consistency in secure persistent memory}. In
  \bibinfo{booktitle}{\emph{MICRO-54: 54th Annual IEEE/ACM International
  Symposium on Microarchitecture}}. \bibinfo{pages}{1227--1240}.
\newblock


\bibitem[Gassend et~al\mbox{.}(2003)]%
        {Gassend_Suh_Clarke_van}
\bibfield{author}{\bibinfo{person}{Blaise Gassend}, \bibinfo{person}{G~Edward
  Suh}, \bibinfo{person}{Dwaine Clarke}, \bibinfo{person}{Marten Van~Dijk},
  {and} \bibinfo{person}{Srinivas Devadas}.} \bibinfo{year}{2003}\natexlab{}.
\newblock \showarticletitle{Caches and hash trees for efficient memory
  integrity verification}. In \bibinfo{booktitle}{\emph{The Ninth International
  Symposium on High-Performance Computer Architecture, 2003. HPCA-9 2003.
  Proceedings.}} IEEE, \bibinfo{pages}{295--306}.
\newblock


\bibitem[Hofmann et~al\mbox{.}(2013)]%
        {hofmann2013inktag}
\bibfield{author}{\bibinfo{person}{Owen~S Hofmann}, \bibinfo{person}{Sangman
  Kim}, \bibinfo{person}{Alan~M Dunn}, \bibinfo{person}{Michael~Z Lee}, {and}
  \bibinfo{person}{Emmett Witchel}.} \bibinfo{year}{2013}\natexlab{}.
\newblock \showarticletitle{Inktag: Secure applications on an untrusted
  operating system}. In \bibinfo{booktitle}{\emph{Proceedings of the eighteenth
  international conference on Architectural support for programming languages
  and operating systems}}. \bibinfo{pages}{265--278}.
\newblock


\bibitem[Huang and Hua(2021)]%
        {Huang_Hua_2021}
\bibfield{author}{\bibinfo{person}{Jianming Huang} {and} \bibinfo{person}{Yu
  Hua}.} \bibinfo{year}{2021}\natexlab{}.
\newblock \showarticletitle{A Write-Friendly and Fast-Recovery Scheme for
  Security Metadata in Non-Volatile Memories}. In
  \bibinfo{booktitle}{\emph{2021 IEEE International Symposium on
  High-Performance Computer Architecture (HPCA)}}.
\newblock
\urldef\tempurl%
\url{https://doi.org/10.1109/hpca51647.2021.00038}
\showDOI{\tempurl}


\bibitem[Lee et~al\mbox{.}(2017)]%
        {inferring}
\bibfield{author}{\bibinfo{person}{Sangho Lee}, \bibinfo{person}{Ming-Wei
  Shih}, \bibinfo{person}{Prasun Gera}, \bibinfo{person}{Taesoo Kim},
  \bibinfo{person}{Hyesoon Kim}, {and} \bibinfo{person}{Marcus Peinado}.}
  \bibinfo{year}{2017}\natexlab{}.
\newblock \showarticletitle{Inferring fine-grained control flow inside
  $\{$SGX$\}$ enclaves with branch shadowing}. In
  \bibinfo{booktitle}{\emph{26th USENIX Security Symposium (USENIX Security
  17)}}. \bibinfo{pages}{557--574}.
\newblock


\bibitem[Lesjak et~al\mbox{.}(2015)]%
        {Alves_2004}
\bibfield{author}{\bibinfo{person}{Christian Lesjak}, \bibinfo{person}{Daniel
  Hein}, {and} \bibinfo{person}{Johannes Winter}.}
  \bibinfo{year}{2015}\natexlab{}.
\newblock \showarticletitle{Hardware-security technologies for industrial IoT:
  TrustZone and security controller}.
\newblock  (\bibinfo{year}{2015}), \bibinfo{pages}{002589--002595}.
\newblock


\bibitem[Lie et~al\mbox{.}(2003)]%
        {lie2003implementing}
\bibfield{author}{\bibinfo{person}{David Lie}, \bibinfo{person}{Chandramohan~A
  Thekkath}, {and} \bibinfo{person}{Mark Horowitz}.}
  \bibinfo{year}{2003}\natexlab{}.
\newblock \showarticletitle{Implementing an untrusted operating system on
  trusted hardware}. In \bibinfo{booktitle}{\emph{Proceedings of the nineteenth
  ACM symposium on Operating systems principles}}. \bibinfo{pages}{178--192}.
\newblock


\bibitem[Limaye and Adegbija(2018)]%
        {spec2017cpu}
\bibfield{author}{\bibinfo{person}{Ankur Limaye} {and} \bibinfo{person}{Tosiron
  Adegbija}.} \bibinfo{year}{2018}\natexlab{}.
\newblock \bibinfo{title}{A workload characterization of the spec cpu2017
  benchmark suite}.
\newblock , \bibinfo{numpages}{149--158}~pages.
\newblock


\bibitem[Liu et~al\mbox{.}(2022)]%
        {liu2022ps}
\bibfield{author}{\bibinfo{person}{Gang Liu}, \bibinfo{person}{Kenli Li},
  \bibinfo{person}{Zheng Xiao}, {and} \bibinfo{person}{Rujia Wang}.}
  \bibinfo{year}{2022}\natexlab{}.
\newblock \showarticletitle{Ps-oram: Efficient crash consistency support for
  oblivious ram on nvm}. In \bibinfo{booktitle}{\emph{Proceedings of the 49th
  Annual International Symposium on Computer Architecture}}.
  \bibinfo{pages}{188--203}.
\newblock


\bibitem[Liu et~al\mbox{.}(2025)]%
        {liu2025hm}
\bibfield{author}{\bibinfo{person}{Gang Liu}, \bibinfo{person}{Zheng Xiao},
  \bibinfo{person}{kenli Li}, {and} \bibinfo{person}{Rujia Wang}.}
  \bibinfo{year}{2025}\natexlab{}.
\newblock \showarticletitle{HM-ORAM: A Lightweight Crash-consistent ORAM
  Framework on Hybrid Memory System}.
\newblock \bibinfo{journal}{\emph{ACM Transactions on Storage}}
  \bibinfo{volume}{21}, \bibinfo{number}{2} (\bibinfo{year}{2025}),
  \bibinfo{pages}{1--28}.
\newblock


\bibitem[Lowe-Power et~al\mbox{.}(2020)]%
        {lowe2021gem5}
\bibfield{author}{\bibinfo{person}{Jason Lowe-Power},
  \bibinfo{person}{Abdul~Mutaal Ahmad}, \bibinfo{person}{Ayaz Akram},
  \bibinfo{person}{Mohammad Alian}, \bibinfo{person}{Rico Amslinger},
  \bibinfo{person}{Matteo Andreozzi}, \bibinfo{person}{Adri{\`a} Armejach},
  \bibinfo{person}{Nils Asmussen}, \bibinfo{person}{Brad Beckmann},
  \bibinfo{person}{Srikant Bharadwaj}, {et~al\mbox{.}}}
  \bibinfo{year}{2020}\natexlab{}.
\newblock \showarticletitle{The gem5 simulator: Version 20.0+}.
\newblock \bibinfo{journal}{\emph{arXiv preprint arXiv:2007.03152}}
  (\bibinfo{year}{2020}).
\newblock


\bibitem[McCune et~al\mbox{.}(2008)]%
        {mccune2008flicker}
\bibfield{author}{\bibinfo{person}{Jonathan~M McCune}, \bibinfo{person}{Bryan~J
  Parno}, \bibinfo{person}{Adrian Perrig}, \bibinfo{person}{Michael~K Reiter},
  {and} \bibinfo{person}{Hiroshi Isozaki}.} \bibinfo{year}{2008}\natexlab{}.
\newblock \showarticletitle{Flicker: An execution infrastructure for TCB
  minimization}. In \bibinfo{booktitle}{\emph{Proceedings of the 3rd ACM
  SIGOPS/EuroSys European Conference on Computer Systems 2008}}.
  \bibinfo{pages}{315--328}.
\newblock


\bibitem[McKeen et~al\mbox{.}(2016a)]%
        {rozas2013intel}
\bibfield{author}{\bibinfo{person}{Frank McKeen}, \bibinfo{person}{Ilya
  Alexandrovich}, \bibinfo{person}{Ittai Anati}, \bibinfo{person}{Dror Caspi},
  \bibinfo{person}{Simon Johnson}, \bibinfo{person}{Rebekah Leslie-Hurd}, {and}
  \bibinfo{person}{Carlos Rozas}.} \bibinfo{year}{2016}\natexlab{a}.
\newblock \bibinfo{title}{Intel{\textregistered} software guard extensions
  (intel{\textregistered} sgx) support for dynamic memory management inside an
  enclave}.
\newblock , \bibinfo{numpages}{9}~pages.
\newblock


\bibitem[McKeen et~al\mbox{.}(2016b)]%
        {mckeen2016intel}
\bibfield{author}{\bibinfo{person}{Frank McKeen}, \bibinfo{person}{Ilya
  Alexandrovich}, \bibinfo{person}{Ittai Anati}, \bibinfo{person}{Dror Caspi},
  \bibinfo{person}{Simon Johnson}, \bibinfo{person}{Rebekah Leslie-Hurd}, {and}
  \bibinfo{person}{Carlos Rozas}.} \bibinfo{year}{2016}\natexlab{b}.
\newblock \showarticletitle{Intel{\textregistered} software guard extensions
  (intel{\textregistered} sgx) support for dynamic memory management inside an
  enclave}.
\newblock In \bibinfo{booktitle}{\emph{Proceedings of the Hardware and
  Architectural Support for Security and Privacy 2016}}. \bibinfo{pages}{1--9}.
\newblock


\bibitem[McKeen et~al\mbox{.}(2013a)]%
        {McKeen2013}
\bibfield{author}{\bibinfo{person}{Frank McKeen}, \bibinfo{person}{Ilya
  Alexandrovich}, \bibinfo{person}{Alex Berenzon}, \bibinfo{person}{Carlos~V
  Rozas}, \bibinfo{person}{Hisham Shafi}, \bibinfo{person}{Vedvyas Shanbhogue},
  {and} \bibinfo{person}{Uday~R Savagaonkar}.}
  \bibinfo{year}{2013}\natexlab{a}.
\newblock \showarticletitle{Innovative instructions and software model for
  isolated execution.}
\newblock \bibinfo{journal}{\emph{Hasp@ isca}} \bibinfo{volume}{10},
  \bibinfo{number}{1}.
\newblock


\bibitem[McKeen et~al\mbox{.}(2013b)]%
        {mckeen2013innovative}
\bibfield{author}{\bibinfo{person}{Frank McKeen}, \bibinfo{person}{Ilya
  Alexandrovich}, \bibinfo{person}{Alex Berenzon}, \bibinfo{person}{Carlos~V
  Rozas}, \bibinfo{person}{Hisham Shafi}, \bibinfo{person}{Vedvyas Shanbhogue},
  {and} \bibinfo{person}{Uday~R Savagaonkar}.}
  \bibinfo{year}{2013}\natexlab{b}.
\newblock \showarticletitle{Innovative instructions and software model for
  isolated execution.}
\newblock \bibinfo{journal}{\emph{Hasp@ isca}} \bibinfo{volume}{10},
  \bibinfo{number}{1} (\bibinfo{year}{2013}).
\newblock


\bibitem[Oleksenko et~al\mbox{.}(2018)]%
        {oleksenko2018varys}
\bibfield{author}{\bibinfo{person}{Oleksii Oleksenko}, \bibinfo{person}{Bohdan
  Trach}, \bibinfo{person}{Robert Krahn}, \bibinfo{person}{Mark Silberstein},
  {and} \bibinfo{person}{Christof Fetzer}.} \bibinfo{year}{2018}\natexlab{}.
\newblock \showarticletitle{Varys: Protecting $\{$SGX$\}$ enclaves from
  practical $\{$Side-Channel$\}$ attacks}. In \bibinfo{booktitle}{\emph{2018
  Usenix Annual Technical Conference (USENIX ATC 18)}}.
  \bibinfo{pages}{227--240}.
\newblock


\bibitem[Orenbach et~al\mbox{.}(2020)]%
        {orenbach2020autarky}
\bibfield{author}{\bibinfo{person}{Meni Orenbach}, \bibinfo{person}{Andrew
  Baumann}, {and} \bibinfo{person}{Mark Silberstein}.}
  \bibinfo{year}{2020}\natexlab{}.
\newblock \showarticletitle{Autarky: Closing controlled channels with
  self-paging enclaves}. In \bibinfo{booktitle}{\emph{Proceedings of the
  Fifteenth European Conference on Computer Systems}}. \bibinfo{pages}{1--16}.
\newblock


\bibitem[Ports and Garfinkel(2008)]%
        {ports2008towards}
\bibfield{author}{\bibinfo{person}{Dan~RK Ports} {and} \bibinfo{person}{Tal
  Garfinkel}.} \bibinfo{year}{2008}\natexlab{}.
\newblock \showarticletitle{Towards Application Security on Untrusted Operating
  Systems.}. In \bibinfo{booktitle}{\emph{HotSec}}.
\newblock


\bibitem[Rogers et~al\mbox{.}(2007)]%
        {Rogers_Chhabra_Prvulovic_Solihin_2007}
\bibfield{author}{\bibinfo{person}{Brian Rogers}, \bibinfo{person}{Siddhartha
  Chhabra}, \bibinfo{person}{Milos Prvulovic}, {and} \bibinfo{person}{Yan
  Solihin}.} \bibinfo{year}{2007}\natexlab{}.
\newblock \showarticletitle{Using address independent seed encryption and
  bonsai merkle trees to make secure processors os-and performance-friendly}.
  In \bibinfo{booktitle}{\emph{40th Annual IEEE/ACM International Symposium on
  Microarchitecture (MICRO 2007)}}. IEEE, \bibinfo{pages}{183--196}.
\newblock


\bibitem[Sasy et~al\mbox{.}(2017)]%
        {zerotrace}
\bibfield{author}{\bibinfo{person}{Sajin Sasy}, \bibinfo{person}{Sergey
  Gorbunov}, {and} \bibinfo{person}{Christopher~W Fletcher}.}
  \bibinfo{year}{2017}\natexlab{}.
\newblock \showarticletitle{ZeroTrace: Oblivious memory primitives from Intel
  SGX}.
\newblock \bibinfo{journal}{\emph{Cryptology ePrint Archive}}.
\newblock


\bibitem[Shadab et~al\mbox{.}(2023)]%
        {Shadab_Zou_Gandham_Awad_Lin_Mingjie}
\bibfield{author}{\bibinfo{person}{Rakin~Muhammad Shadab}, \bibinfo{person}{Yu
  Zou}, \bibinfo{person}{Sanjay Gandham}, \bibinfo{person}{Amro Awad}, {and}
  \bibinfo{person}{Mingjie Lin}.} \bibinfo{year}{2023}\natexlab{}.
\newblock \showarticletitle{A secure computing system with hardware-efficient
  lazy bonsai merkle tree for fpga-attached embedded memory}.
\newblock \bibinfo{journal}{\emph{IEEE Transactions on Dependable and Secure
  Computing}} \bibinfo{volume}{21}, \bibinfo{number}{4} (\bibinfo{year}{2023}),
  \bibinfo{pages}{3262--3279}.
\newblock


\bibitem[Shih et~al\mbox{.}(2017)]%
        {shih2017t}
\bibfield{author}{\bibinfo{person}{Ming-Wei Shih}, \bibinfo{person}{Sangho
  Lee}, \bibinfo{person}{Taesoo Kim}, {and} \bibinfo{person}{Marcus Peinado}.}
  \bibinfo{year}{2017}\natexlab{}.
\newblock \showarticletitle{T-SGX: Eradicating Controlled-Channel Attacks
  Against Enclave Programs.}. In \bibinfo{booktitle}{\emph{NDSS}},
  Vol.~\bibinfo{volume}{6}. \bibinfo{pages}{15--43}.
\newblock


\bibitem[Shinde et~al\mbox{.}(2016)]%
        {shinde2016preventing}
\bibfield{author}{\bibinfo{person}{Shweta Shinde}, \bibinfo{person}{Zheng~Leong
  Chua}, \bibinfo{person}{Viswesh Narayanan}, {and} \bibinfo{person}{Prateek
  Saxena}.} \bibinfo{year}{2016}\natexlab{}.
\newblock \showarticletitle{Preventing page faults from telling your secrets}.
  In \bibinfo{booktitle}{\emph{Proceedings of the 11th ACM on Asia Conference
  on Computer and Communications Security}}. \bibinfo{pages}{317--328}.
\newblock


\bibitem[Sinha et~al\mbox{.}(2017)]%
        {compiler}
\bibfield{author}{\bibinfo{person}{Rohit Sinha}, \bibinfo{person}{Sriram
  Rajamani}, {and} \bibinfo{person}{Sanjit~A Seshia}.}
  \bibinfo{year}{2017}\natexlab{}.
\newblock \showarticletitle{A compiler and verifier for page access oblivious
  computation}. In \bibinfo{booktitle}{\emph{Proceedings of the 2017 11th Joint
  Meeting on Foundations of Software Engineering}}. \bibinfo{pages}{649--660}.
\newblock


\bibitem[Ta-Min et~al\mbox{.}(2006)]%
        {ta2006splitting}
\bibfield{author}{\bibinfo{person}{Richard Ta-Min}, \bibinfo{person}{Lionel
  Litty}, {and} \bibinfo{person}{David Lie}.} \bibinfo{year}{2006}\natexlab{}.
\newblock \showarticletitle{Splitting interfaces: Making trust between
  applications and operating systems configurable}. In
  \bibinfo{booktitle}{\emph{Proceedings of the 7th symposium on Operating
  systems design and implementation}}. \bibinfo{pages}{279--292}.
\newblock


\bibitem[Wichelmann et~al\mbox{.}(2024)]%
        {wichelmann2024obelix}
\bibfield{author}{\bibinfo{person}{Jan Wichelmann}, \bibinfo{person}{Anja
  Rabich}, \bibinfo{person}{Anna P{\"a}tschke}, {and} \bibinfo{person}{Thomas
  Eisenbarth}.} \bibinfo{year}{2024}\natexlab{}.
\newblock \showarticletitle{Obelix: Mitigating side-channels through dynamic
  obfuscation}. In \bibinfo{booktitle}{\emph{2024 IEEE Symposium on Security
  and Privacy (SP)}}. IEEE, \bibinfo{pages}{4182--4199}.
\newblock


\bibitem[Xu et~al\mbox{.}(2015)]%
        {Controlled-Channel-Attacks}
\bibfield{author}{\bibinfo{person}{Yuanzhong Xu}, \bibinfo{person}{Weidong
  Cui}, {and} \bibinfo{person}{Marcus Peinado}.}
  \bibinfo{year}{2015}\natexlab{}.
\newblock \showarticletitle{Controlled-channel attacks: Deterministic side
  channels for untrusted operating systems}. In \bibinfo{booktitle}{\emph{2015
  IEEE Symposium on Security and Privacy}}. IEEE, \bibinfo{pages}{640--656}.
\newblock


\bibitem[Yadalam et~al\mbox{.}(2020)]%
        {yadalam2020sgxl}
\bibfield{author}{\bibinfo{person}{Sujay Yadalam}, \bibinfo{person}{Vinod
  Ganapathy}, {and} \bibinfo{person}{Arkaprava Basu}.}
  \bibinfo{year}{2020}\natexlab{}.
\newblock \showarticletitle{SGXL: Security and Performance for Enclaves Using
  Large Pages}.
\newblock \bibinfo{journal}{\emph{ACM Transactions on Architecture and Code
  Optimization (TACO)}} \bibinfo{volume}{18}, \bibinfo{number}{1}
  (\bibinfo{year}{2020}), \bibinfo{pages}{1--25}.
\newblock


\bibitem[Yang and Shin(2008)]%
        {yang2008using}
\bibfield{author}{\bibinfo{person}{Jisoo Yang} {and} \bibinfo{person}{Kang~G
  Shin}.} \bibinfo{year}{2008}\natexlab{}.
\newblock \showarticletitle{Using hypervisor to provide data secrecy for user
  applications on a per-page basis}. In \bibinfo{booktitle}{\emph{Proceedings
  of the fourth ACM SIGPLAN/SIGOPS international conference on Virtual
  execution environments}}. \bibinfo{pages}{71--80}.
\newblock


\bibitem[Zhang et~al\mbox{.}(2011)]%
        {zhang2011cloudvisor}
\bibfield{author}{\bibinfo{person}{Fengzhe Zhang}, \bibinfo{person}{Jin Chen},
  \bibinfo{person}{Haibo Chen}, {and} \bibinfo{person}{Binyu Zang}.}
  \bibinfo{year}{2011}\natexlab{}.
\newblock \showarticletitle{Cloudvisor: retrofitting protection of virtual
  machines in multi-tenant cloud with nested virtualization}. In
  \bibinfo{booktitle}{\emph{Proceedings of the twenty-third acm symposium on
  operating systems principles}}. \bibinfo{pages}{203--216}.
\newblock


\bibitem[Zhang et~al\mbox{.}(2020)]%
        {zhang2020klotski}
\bibfield{author}{\bibinfo{person}{Pan Zhang}, \bibinfo{person}{Chengyu Song},
  \bibinfo{person}{Heng Yin}, \bibinfo{person}{Deqing Zou},
  \bibinfo{person}{Elaine Shi}, {and} \bibinfo{person}{Hai Jin}.}
  \bibinfo{year}{2020}\natexlab{}.
\newblock \showarticletitle{Klotski: Efficient obfuscated execution against
  controlled-channel attacks}. In \bibinfo{booktitle}{\emph{Proceedings of the
  Twenty-Fifth International Conference on Architectural Support for
  Programming Languages and Operating Systems}}. \bibinfo{pages}{1263--1276}.
\newblock


\bibitem[Zhao et~al\mbox{.}(2013)]%
        {zhao2013kiln}
\bibfield{author}{\bibinfo{person}{Jishen Zhao}, \bibinfo{person}{Sheng Li},
  \bibinfo{person}{Doe~Hyun Yoon}, \bibinfo{person}{Yuan Xie}, {and}
  \bibinfo{person}{Norman~P Jouppi}.} \bibinfo{year}{2013}\natexlab{}.
\newblock \showarticletitle{Kiln: Closing the performance gap between systems
  with and without persistence support}. In
  \bibinfo{booktitle}{\emph{Proceedings of the 46th Annual IEEE/ACM
  International Symposium on Microarchitecture}}. \bibinfo{pages}{421--432}.
\newblock


\end{thebibliography}


\end{document}